\documentclass[preprint,authoryear,12pt]{elsarticle}
\usepackage{rotate, color}
\usepackage{amsmath, amssymb, theorem}
\usepackage{setspace}
\usepackage{graphicx}
\usepackage{morefloats}
\usepackage{subfig}
\usepackage{pdflscape}
\usepackage{pgfplots}
\usepackage{multirow}
\usepackage{booktabs}
\usepackage{url}
\usepackage{appendix}
\usepackage{threeparttable}
\usepackage{lineno,hyperref}
\modulolinenumbers[5]
\newcommand{\be}{\begin{equation}}
\newcommand{\ee}{\end{equation}}
\newcommand{\bean}{\begin{eqnarray*}}
\newcommand{\eean}{\end{eqnarray*}}

\newtheorem{remark}{Remark}

\theoremstyle{definition}

\theoremstyle{remark}

\definecolor{purple}{rgb}{0.7,0.0,0.8}

\journal{Computational Statistics \& Data Analysis}
\makeatletter
\def\ps@pprintTitle{%
  \let\@oddhead\@empty
  \let\@evenhead\@empty
  \let\@oddfoot\@empty
  \let\@evenfoot\@oddfoot
}
\makeatother

\begin{document}

\begin{frontmatter}

\title{Active learning procedure via sequential experimental design and uncertainty sampling}

\author[a1]{Jing Wang}
\author[a1]{Eunsik Park}
\address[a1]{Department of Statistics, Chonnam National University, Gwangju 500-757, Korea}
% \fntext[myfootnote]{Since 1880.}
\author[a2]{Yuan-chin Ivan Chang}
\address[a2]{Institute of Statistical Science, Academia Sinica, Taipei 11529, Taiwan}
\fntext[]{\text{Correspondence to E. Park, espark02@gmail.com and Y-c Ivan Chang, ycchang@sinica.edu.tw} \hfill}

\color{black}
\begin{abstract}
Classification is an important task in many fields including biomedical research and machine learning.  Traditionally, a classification rule is constructed based a bunch of labeled data.  Recently, due to technological innovation and automatic data collection schemes, we easily encounter with data sets containing large amounts of unlabeled samples.  Because to label each of them is usually costly and inefficient,  how to  utilize these unlabeled data in a classifier construction process becomes an important problem.  In machine learning literature, active learning or semi-supervised learning are popular concepts discussed under this situation, where classification algorithms recruit new unlabeled subjects sequentially based on the information learned from previous stages of its learning process, and these new subjects  are then labeled and included as new training samples.  From a statistical aspect,  these methods can be recognized as a hybrid of the sequential design and stochastic approximation procedure.  In this paper, we study sequential learning procedures for building efficient and effective classifiers, where only the selected subjects are labeled and included in its learning stage.  The proposed algorithm combines the ideas of Bayesian sequential optimal design and uncertainty sampling. Computational issues of the algorithm are discussed. Numerical results using both synthesized data and real examples are reported.
\end{abstract}


\end{frontmatter}

% \linenumbers
\color{black}
\section{Introduction}
Classification is an important task in many fields including biomedical research, engineering, sociology and many others.  How to construct a classification rule based on a labeled data set is a classical statistical problem. In machine learning literature, there are several types of learning problems discussed, and depending on how labeled subjects are included into a learning process, they are usually termed as supervised, unsupervised or semi-supervised learning  \citep{S2000, SB2010}.  Recently, due to technical innovation, ``big data'' becomes a buzz phrase in  many fields, and we now often encounter with data sets that have huge amount of unlabeled data. Hence, how to utilize these unlabeled data efficiently to construct a classification rule becomes an important problem.
Because to label each unlabeled subject is usually costly and inefficient, a common approach is active learning \citep[see, for example,][]{CGJ1996, YBT2006}.  This type of a leaning process will only inquire the label information for the ``selected'' subjects, which are usually chosen based on the information learned in the previous learning stages, and then include the newly labeled subjects into its training stage. A learning process will usually go on until a prefixed criterion is reached, such as a prefixed total number of labeled subjects to be used in the training stage.

Moreover, because in an active learning process, subjects are dynamically and sequentially selected, labeled and then added to the training set, this process is naturally related to sequential experimental designs in Statistics, where a new observation/experiment is conducted at some particular design points selected according to the information obtained using the data gathered up to current stage.  Since data are observed adaptively, this type of methods are also related to the stochastic approximation process, which was first discussed in \citet{RM1951}.  Their original procedure is called Robbins-Monro (RM) procedure and can be viewed as a stochastic version of  Newton-Raphson method for nonlinear root-finding problems.  Following \citet{RM1951}, sequential design methods have been intensively studied, and there are even more papers discussed different modifications of RM procedure and their corresponding convergence rates. Recently, \citet{Jo2004} further modified RM procedure to improve on its efficiency.  This type of procedures is nonparametric in the sense that no parametric model assumption is presumed.

However, RM procedure can also be derived from a parametric form. For example, using the maximum likelihood estimate (MLE) of a logistic model, \citet{Wu1985} proposed a logit-MLE method for binary data that  uses the currently available labeled data to fit a logistic model, and then select the next input with the desired probability based on the fitted logistic model.
Because a classification rule construction under active learning framework can be formulated as a problem of estimating the threshold boundary between two groups, which can usually be defined using a probability quantile, it can also be viewed as a stochastic root-finding procedure described above.
Moreover, logistic models are commonly used models in binary classification problems, and the properties of sequential estimation for generalized linear model (GLM) under general adaptive designs are well studied \citep{chang2001, Zacks08}.
Hence, it is natural to construct a binary classification rule, sequentially and adaptively, by putting all these ingredients together.  An active learning algorithm developed in \citet{Deng2009}, which combines the logic-MLE of \citet{Wu1985} and D-optimal design is a successful example.
\textcolor{black}{This kind of a method depends on the properties of MLE.}
Although, the existence and uniqueness of MLE can be achieved after quite a few initial observations \citep{S1981}, it may still suffer from severe bias, when sample size is small, which usually results in an inefficient learning process.
In modern literature,  \citet{joseph2007adaptive} developed a Bayesian extension of Wu's approach, where they used the maximum a posterior (MAP) estimates of the  parameters of a logistic model rather than MLEs. \citet{DS2008} suggested a new sequential experimental design for GLM, where observations are selected sequentially based on a Bayesian D-optimality criterion and Bayesian estimates of model parameters.  These methods motivate us to study a novel modification of \citet{Deng2009}.

As in conventional regression analysis, it is well-known that when the number of dimensionality of the unknown vector of parameters becomes large, the estimated information of it will be very unstable. Because active learning processes usually rely such kind of information, the unstable estimates of parameters will also affect the learning process.
In the real example studied in \citet{Deng2009}, those two variables are selected based on experts' opinions.  However, this situation is rare and there are usually more variables considered for a real example. Thus, how to stabilize a learning process in high dimensional case is difficult and important.  In this paper, we focus on the higher dimensional data sets.     A Bayesian sequential design is used and the related computational issues are discussed.   In addition, for practical usages, we also study the effects of using different sizes of labeled data sets as an initial training set of an active learning process. As to the subject selection during a process, the major difference between a sequential design and an active learning process is that with sequential design, an experiment will be conducted at the selected points, while in active learning processes with existent unlabeled data, we can only select points near the theoretical ones from an existent data set.  Hence, how to select the next point based on the available information plays a key role in an active learning process. \citet{Deng2009} aimed at shortening the distance between the estimated boundary and the true one, such that their subject selection scheme heavily depends on the initial model assumption.  In practice, the form of true model is usually unknown. Hence, in order to diminish the effect of model assumptions, we adopt a different design point selection scheme.  The advantage of the proposed method will be discussed from both theoretical and practical aspects.

The rest of this paper is organized as follows. In Section \ref{sec: Method}, we first review the active learning algorithm \citet{Deng2009}, and then discuss the proposed algorithm and some modifications. Simulation results and numerical studies with real data sets are presented in Sections \ref{sec: Sim} and \ref{sec: Real}, respectively. Section \ref{sec: Discu} is a summarization.  Technical details are given in Appendix.

\color{black}
\section{Methodology}
\label{sec: Method}

\subsection{Model and Parameter Estimation}
\label{sec: AL}

Let $ \mathbf{x} =(x_{1}, \ldots ,x_{p})^{T} $ be the explanatory vector of subjects and variable $ Y=1 $ or $ Y=0 $ denotes the category a subject belonging to.
Suppose that $ P(Y=1|\mathbf{x})=F(\mathbf{x})$ be the probability model of $ Y=1 $ given $ \mathbf {x} $. 
Assume further  that each variable has a positive relationship with the response;  that is, for larger value of $ x_j $, the higher the probability of $ Y = 1 $. 
Then \citet{Deng2009} assumed that $ F(\mathbf {x}) $ had a parametric form
\begin{align}\label{model-1}{
  F(\mathbf{x}|\boldsymbol\theta) = \frac{ e^{(z-\mu)/\sigma} }{ 1+e^{(z-\mu)/\sigma} },
  }
\end{align}
where $ z = \sum_{i=1}^p w_{i}x_i $, $ 0 < w_i < 1 $ for each $i$, and $ \sum_{i=1}^p w_i = 1 $.
Let $ \boldsymbol\theta = (\mu,\sigma,w_1 \ldots ,w_{p-1})^{T} $ be a vector of $p+1$ parameters.
Then following (\ref{model-1}), for a given $\mathbf{x}$, $ Y $ is a Bernoulli random variable with  mean  $ E(Y|\mathbf{x})=F(\mathbf{x}|\boldsymbol\theta) $.
 Model (\ref{model-1}) can be re-written as a conventional logistic regression model:
\begin{align}\label{model-2}{
  F(\mathbf{x}|\boldsymbol\beta) = \frac{e^{\tilde{\mathbf{x}}^{T} \boldsymbol \beta} }{1 + e^{\tilde{\mathbf{x}}^{T} \boldsymbol \beta}},
  }
\end{align}
where $ \tilde{\mathbf{x}}^{T} = (1, \mathbf{x}^{T}) $ and $ \boldsymbol \beta = ( -\mu / \sigma,~w_1 / \sigma,~\ldots~,~w_p / \sigma )^{T} $.
The Fisher information matrix of $\boldsymbol \beta$ with a set of design points $ d = \{ \mathbf {x}_1, \ldots, \mathbf {x}_n \} $ is
\begin{align}\label{model-3}{
  \mathbf{I}(\boldsymbol\beta;d)= \mathbf{X}^{T}\mathbf{W}\mathbf{X},
  }
\end{align}
where $ \mathbf {X} $ is the regression matrix with $i$th row, $i=1, \ldots, p$ equal to $ (1, x_{i1}, \ldots, x_{ip}) $ and $ \mathbf {W} $ is a diagonal matrix with $ w_{ii} = F(\mathbf{x}_i|\boldsymbol\beta)\left[1-F(\mathbf{x}_i|\boldsymbol\beta)\right] $, $i = 1,...,p$.  It is clear that this information matrix is non-linear in $\boldsymbol \beta$ and depends on the unknown $\boldsymbol \beta$ only through $ \mathbf {W} $.

Suppose that $ (\mathbf {x}_1,Y_1),\cdots, (\mathbf {x}_n,Y_n) $ are  observed labeled data of size $n$. Using this training set,  we obtain an  \textcolor{black}{MAP estimates of  both $ \boldsymbol{\theta }$ and $\boldsymbol \beta$ and let $\boldsymbol{ \hat\theta }_n = ( \hat{\mu}_n, \hat{\sigma}_n, \hat{w}_{1,n}, \cdots, \hat{w}_{p-1,n})^{T} $ and $ \boldsymbol{ \hat\beta }_n = ( -\hat{\mu}_n / \hat{\sigma}_n,~\hat{w}_{1,n} / \hat{\sigma}_n,~\cdots~,~\hat{w}_{p,n} / \hat{\sigma}_n )^{T} $ denote these two estimates.}
%(The details will be given in Section \ref{sec: Algorithm}.)
Using the current estimates of parameters, the classification rule based on the estimate of $F$ becomes
\begin{align}\label{model-4}{
  \begin{cases}
  \hat F_n ( \mathbf{x}|\boldsymbol{ \hat\beta }_n ) > \textcolor{black}{\gamma},  & \mbox{decide }\mbox{ Y=1 },\\
  \hat F_n ( \mathbf{x}|\boldsymbol{ \hat\beta }_n ) \leqslant \textcolor{black}{\gamma},  & \mbox{decide }\mbox{ Y=0 }
  \end{cases}
  }
\end{align}
with an estimated boundary
\begin{align}\label{model-5}{
  \hat{l}_{n}(\mathbf{x}) = \{ \mathbf{x} =(x_{1}, \cdots ,x_{p})^{T}: \hat F_n ( \mathbf{x}|\boldsymbol{ \hat\beta }_n ) = \omega \},
  }
\end{align}
where $\textcolor{black}{\gamma=0.5}$ when there is no extra information, such as $P(Y=1)$ available.
(In general, the cutting point for a logistic classification function is 0.5.  However,  when there is a prior information about the event, such as prevalence rate in epidemiology study, the cutting point will usually be adjusted accordingly. This will be discussed later.)
Therefore,  the active learning problem under this set up becomes how to recruit a set of training subjects efficiently such that when a learning process is stopped, the final classification function $\hat F_n$ will have good prediction power.

\color{black}
\subsection{Subject Selection}
\label{sec:selection}
Intuitively, in order to have an efficient learning process, we should learn the most uncertain subjects first, because to do it this way may most improve a classifier. Thus,  when using a probabilistic learning model in an active learning framework,  the most commonly used query for getting new data is the uncertainty sampling \citep{SB2010}, where an active learner will query the label information of instance whose class membership is least certain.
For a binary classification problem, this simply means to query the instance whose membership probability is closest to $ 0.5 $ \citep{LG1994, lewis1994heterogenous}.
Thus, in a binary classification case the uncertainty is usually measured by
\begin{align}\label{model-6}
   &\mathbf{d}(\mathbf{x})= \left| \hat F_n ( \mathbf{x}|\boldsymbol{ \hat\beta }_n ) - \omega \right|,
\end{align}
where $\omega = 0.5$.  (Note that in \citet{Deng2009}, they used only one parameter for measuring the uncertainty and adjusting the cutting point, and said that this parameter can be data dependent.  However, our numerical studies show that for our method, using two different parameters for measuring uncertainty and adjusting cutting point, separately, will usually perform better. Regarding this phenomenon, more discussions, based on statistical decision theory viewpoints, are given in Section \ref{sec:uneven}.)

Let $ \mathbf{U} $ be the unlabeled data set.  Then rank points in $ \mathbf{U} $ in ascending order based on  (\ref{model-6}), and an active learning procedure will choose the top ranked point as follows:
\begin{align}
\label{model-7}{
   \mathbf{x}_{n+1}= \arg \min_{\mathbf {x}\in \mathbf{U}} \mathbf{d}(\mathbf{x}).
  }
\end{align}
That is, to choose the one with an estimated probability closest to 0.5 as the next point to be labeled.
Because in high dimensional cases, there may be a lot of points that have the same or similar $\mathbf{d}(\mathbf{x})$, we choose top $ k_n $ points as candidates first, where $ k_n $, in our method, is decided by a local D-efficiency method using a locally optimal design discussed in \citet{woods06} and \citet{DS2008}.   
(For the details of this method, please refer to their original papers.) 
As mentioned in \citet{Deng2009}, to use  (\ref{model-7})  as the only criterion cannot provide good estimates of model parameters, and the method of optimal design can be a good supplement to this disadvantage.  Thus, let  $ C=\{\tilde{\mathbf {x}}_1, \cdots, \tilde{\mathbf {x}}_{k_n} \} $ be the set of candidate points that are screen out using (\ref{model-7}).  We then access these candidates further with some optimal experimental design criterions. 

\begin{figure}[]
          \centering
            \subfloat[]{
	            \begin{minipage}[t]{0.5\linewidth}
	            \includegraphics[width=7cm,height=7cm]{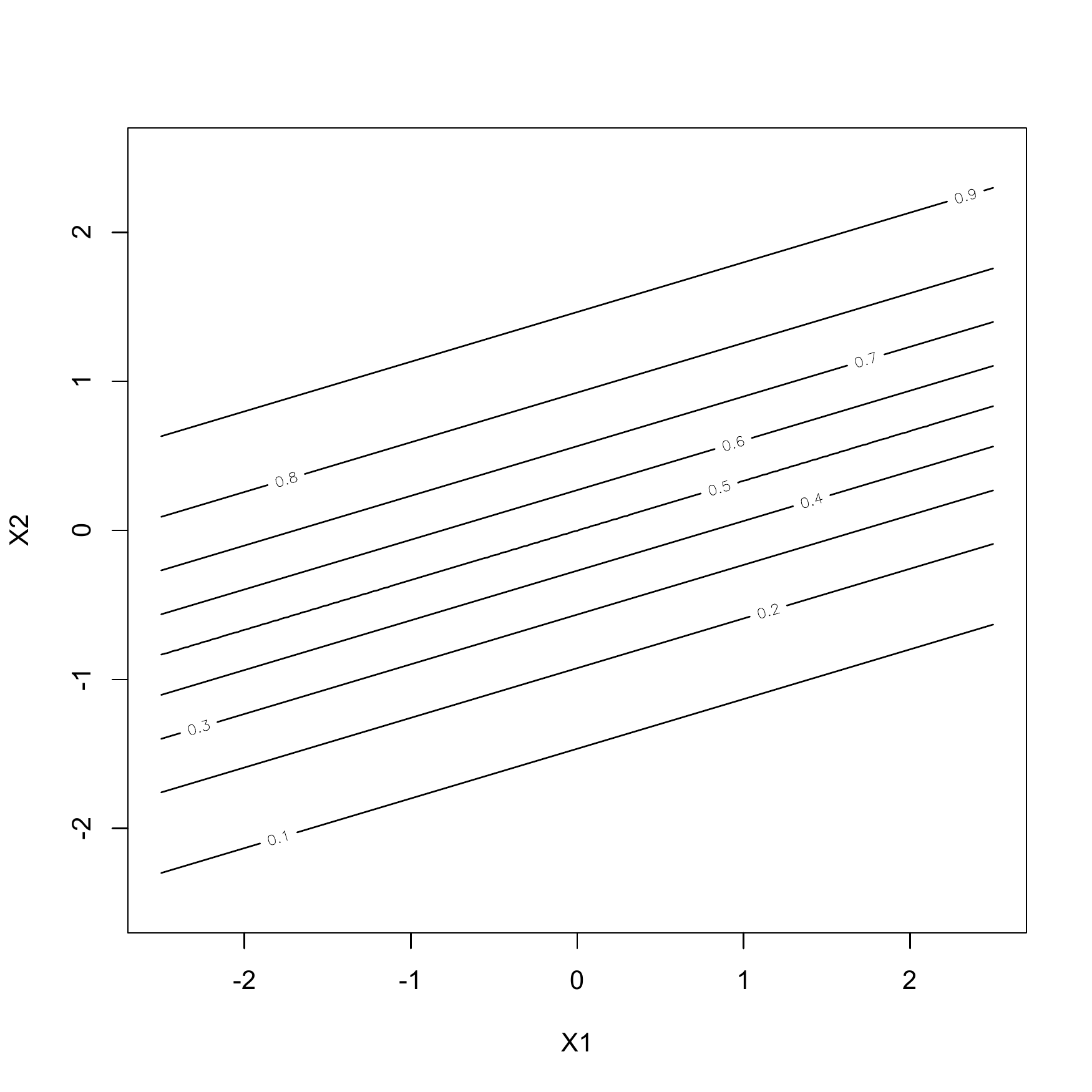}
	            \centering
	            %\caption{(a)}
	            %\label{fig:notification-sys:a}
	            \end{minipage}
	        }
            \subfloat[]{
	            \begin{minipage}[t]{0.5\linewidth}
	            \centering
	            \includegraphics[width=7cm,height=7cm]{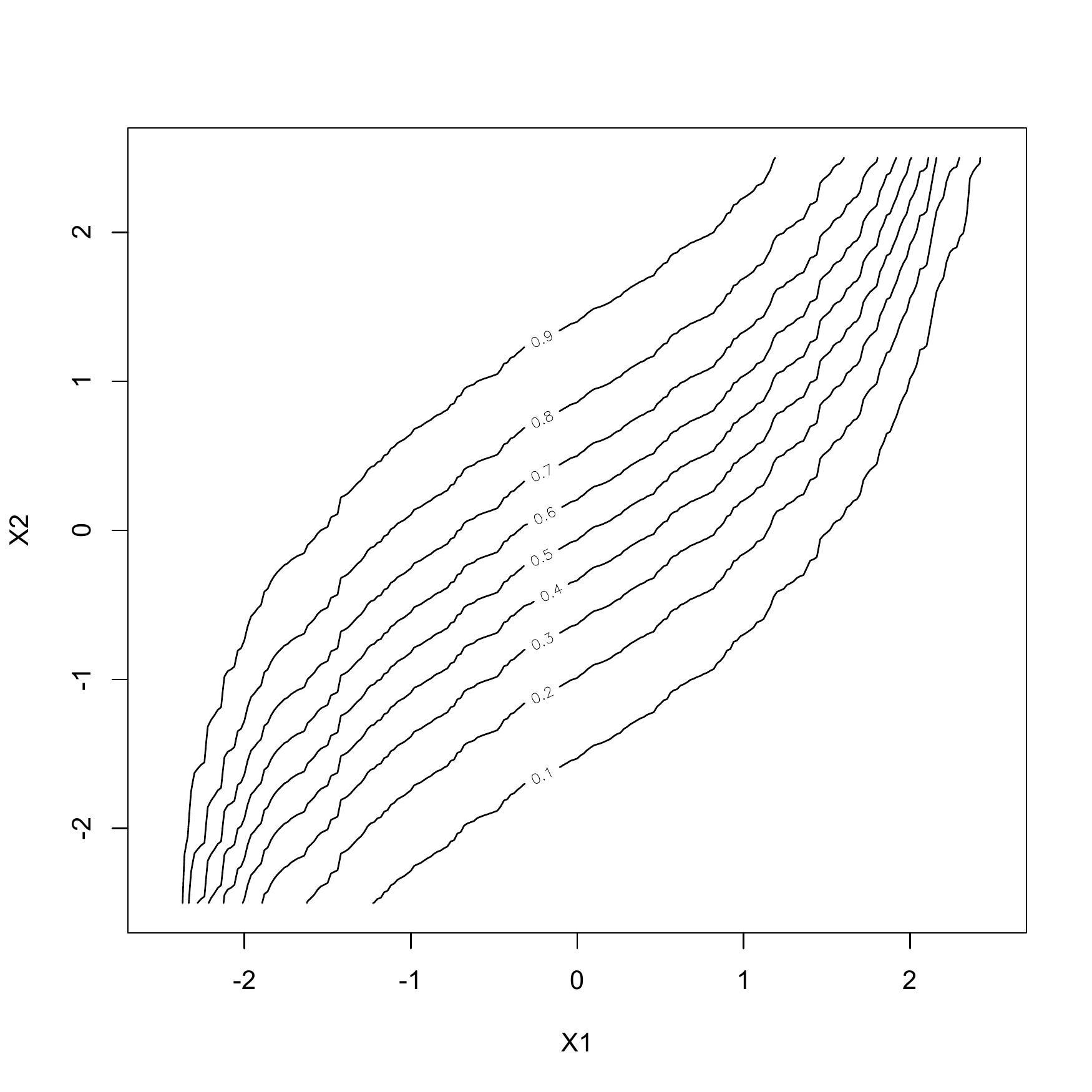}
	            %\caption{(b)}
	           % \label{fig:notification-sys:b}
	            \end{minipage}
            }
          \caption{Contour plot of probabilities produced by two models: (a) When the true model are linear in $X_1$ and $X_2$; (b) When the true model is linear in variables $X_1$ and $X_2$, but has a slight random perturbation. That is, in this case, the linear model is an approximation.}
          \label{Fig:contour}
\end{figure}

\color{black}
One of the major differences between our method and the one in \citet{Deng2009} is that we use an uncertainty sampling method instead of distance based scheme to select the candidate set.  The effect of uncertainty sampling becomes obvious when the difference between the sample sizes of two groups is large.  This situation happens very often in those problems that aim for detecting a set of rare subjects within a large data set, or when the population sizes of two groups are uneven. When the true model is exactly linear and the variables for this model are completely known, these two methods  are the same.  However, in practice, the form of the true model and the variables involved in it are usually unknown.  For instance, the example discussed in  \citet{Deng2009},  those two variables used in their model are selected from a large number of variables by experts, and in fact the true model may involve other variables.   When a model is an approximation with some leftover random errors, then the candidate set defined by a Euclidean distance-based method will be very different from the one obtained using a uncertainty measure.  This situation can be easily illustrated using Figure \ref{Fig:contour}, where Figure \ref{Fig:contour}(a) is the probability contour plot when the true model is linear, and Figure \ref{Fig:contour}(b) is a contour plot of the probabilities for the same linear model plus a small nonlinear error term.   That is, when some perturbation exists, the contour lines can no longer be parallel.  Thus, to use a perpendicular distance to find a candidate set, as that in \citet{Deng2009} cannot be the best choice.   That is the reason why we use an uncertainty sampling scheme to define a candidate set first, then use a (Bayesian) D-optimal design method to screen out the best subject for parameter estimation.   Moreover, when the number of dimensionality becomes larger, the computation of the determinant of a Fisher information matrix is difficult; especially when the size of labeled data is small, the information matrix will be either singular or nearly singular, which provides less information for designs.  Thus, we adopt a Bayesian D-optimal design instead, which will stabilize the beginning stages of a learning process.
 
\color{black}
\paragraph{Computation of Fisher information matrix}

It is known that an active learning is a sequential process, each learning stage heavily relies on the information obtained from its predecessors. Naturally, an unstable initial stage will make the process inefficient and even result in a bias classification rule.
Hence, in order to have a stable learning process, especially in higher dimensional data cases, we adopt a Bayesian D-optimal design instead \citep[see, for example,][]{CL1989, firth1997bayesian}, which is an extension of the original D-optimality by replacing the determinant of Fisher information $\phi_{l}(d)$ with
\begin{align}\label{model-8}{
  \phi(d) \equiv E_{\boldsymbol\beta}\{ \log( |\mathbf{I}(\boldsymbol\beta;d)| ) \} = \int \log ( |\mathbf{I}(\boldsymbol\beta;d)| )\, d \pi(\boldsymbol\beta),
  }
\end{align}
where $ \pi(\boldsymbol\beta) $  denotes the prior distribution for $\boldsymbol\beta$, and the expectation $ E_{\boldsymbol\beta} (\cdot) $ is with  respect to this prior distribution.
%Equation (\ref{model-8}) was also used in \citet{DS2008}. 
Because to compute the integration in (\ref{model-8}) is not trivial, especially when the dimension of $\boldsymbol\beta$ is high.  
This time-consuming step has to be repeated at each stage in an active learning process, hence any simplification of it will be beneficial.   
For this purpose, instead of the exact value of $ \phi(d) $, \citet{DS2008} proposed using an approximation to (\ref{model-8})   below:
\begin{align}\label{model-9}{
  \phi_{1}(d) = \sum_{u=1}^{M}r_{u} \log (|\mathbf{I}(\boldsymbol\beta_{u};d)|),
  }
\end{align}
where $ r_{u} $ are weights obtained with a Monte-Carlo method (see Remark \ref{re:weight}).  A new subject in the candidate set $ C $ that maximizes $\phi_{1}(d) $ is selected.

\begin{remark}\label{re:weight}
The weights $ r_{u} $'s are computed using simple Monte-Carlo method \citep[see][]{niederreiter1988low}. We first generate a large number of points, say M,  from the prior $ \pi(\boldsymbol\beta) $, and denote them as $ \boldsymbol\beta_1,\cdots, \boldsymbol\beta_M $.  Let $ M $ be large enough to represent the prior distribution. 
Then for a vector $ \boldsymbol{\beta}_u $, $u = 1, \ldots, M$, and observations $ Y_1,\cdots, Y_k $ taken at $ \mathbf {x}_1, \cdots, \mathbf {x}_k $, the likelihood is $$ L(\boldsymbol{\beta}_{u})= \prod_{i=1}^k \left[\frac{\exp(\tilde{\mathbf{x}}^{T}_{i}\boldsymbol{\beta}_u)}{1+\exp(\tilde{\mathbf{x}}^{T}_{i}\boldsymbol{\beta}_u)}\right]^{Y_i} \left[\frac{1}{1+\exp(\tilde{\mathbf{x}}^{T}_{i}\boldsymbol{\beta}_u)} \right]^{1-Y_i}. $$ Normalizing the likelihood across the samples, we have weights $ r_u = L(\boldsymbol{\beta}_u)/ \sum_{v=1}^M L(\boldsymbol{\beta}_v) $.
Hence, at each stage of the experiment,  the likelihood for $\boldsymbol{\beta}$ can be rapidly computed.  
\end{remark}

\begin{remark}
In \citet{Deng2009}, they used a local D-optimality criterion to access the unlabeled data in their candidate set (with a prefixed number) and select a subject that maximizes  the determinant of the Fisher information matrix  for $ \boldsymbol \beta $, $ \phi_{l}(d) \equiv |\mathbf{I}(\boldsymbol\beta;d)| $.
This new subject will then be labeled by experts and included to the learning process.
It is clear that the determinant $\phi_{l}(d)$ is numerically unstable when the number of design points is small and the number of dimensionality of $\mathbf x$ is large.
When the information matrix is singular or even just nearly singular,  it is hard to provide useful information for selecting next design points.  Thus, a Bayesian D-optimal design is a good alternative.
\end{remark}

\color{black}
\subsection{The proposed learning algorithm}
\label{sec: Algorithm}

Let $ n_0\geq 0 $ be labeled data points at the initial stage.
Then the proposed algorithm consists of the following steps:
\begin{itemize}

\item[S1.] Compute $ \boldsymbol{\hat\beta}_n $ ---  the posterior estimate of $ \boldsymbol{\beta}$ with the currently available labeled data (When $n_0=0$, we will use the prior median instead);%\todo[inline]{need to explain what to do if $n_0=0$.}

\item[S2.] Rank the unlabeled data points in $ \mathbf{U} $ based on Equation (\ref{model-6}). If the estimated posterior probabilities for all points are equal to either 0 or 1, then stop iteration and use current estimated  $\hat F$ as the final classifier; otherwise, go to S3;

\item[S3.] Create the candidate set $ C $ with the top $ k_n $ points based on the ranks in S2, where $ k_n $ is determined based on the local D-efficiency \citep[see][]{woods06, DS2008};

\item[S4.] Select a new unlabeled point from the set $ C $ according to following criteria:
   \begin{itemize}
    \item[(i)] If the design points up to current stage form a nonsingular information matrix of $\boldsymbol \beta$, then choose the next point that maximizes $ \phi_{1} $ in  (\ref{model-9}); that is,
     \begin{align}\label{model-13}{
       \mathbf {x}_{n+1}= \arg \max_{\mathbf {x}\in C}\phi_{1}(\boldsymbol{\hat\beta}_n;\mathbf {x}_{1},\cdots,\mathbf {x}_{n},\mathbf {x}).
       }
     \end{align}
   \item[(ii)]  If the information matrix is singular, then select the next point from $C$ that maximizes $\phi_{1}$ based on the cumulated $n$ points, $ k_n $-augmentation and the candidate point. That is,
\begin{align}\label{model-14}{
       \mathbf {x}_{n+1}= \arg \max_{\mathbf {x}\in C}\phi_{1}(\boldsymbol{\hat\beta}_n;\mathbf {x}_{1},\cdots,\mathbf {x}_{n},\tilde{\mathbf {x}}_1, \cdots, \tilde{\mathbf {x}}_{k_n},\mathbf {x}).
       }
     \end{align}
     \end{itemize}
\end{itemize}
We consider the case with $\dim(\mathbf x)= p \geq 2 $, so a Dirichlet distribution is a reasonable prior for $ \mathbf{w} = (w_{1}, \cdots, w_{p})^{T} $.  Hence, the following priors are used:
\begin{align}\label{model-15}
\begin{split}
 &       \mu   ~ \sim ~ \mathrm{N}( \mu_0, \sigma^{2}_{\mu}), ~~ \sigma ~ \sim ~ \mathrm{Exponential}(\sigma_0),   \\
 &  \mathbf{w} ~ \sim ~ \mathrm{Dir}(\alpha), ~ \mathrm{where} ~ \alpha = (\alpha_1, \cdots, \alpha_p)^{T}.        \\
\end{split}
\end{align}
Assume that $ \mu $, $ \sigma $ and $ \mathbf{w} $ are mutually independent, then the posterior distribution of $\boldsymbol \theta$, based on the labeled data points $ (\mathbf {x}_1,Y_1),\cdots, (\mathbf {x}_n,Y_n) $ is
\begin{align}\label{model-16}
 \begin{split}
  f(\boldsymbol\theta|\mathbf{Y}) ~ \varpropto ~ & \prod_{i=1}^n \left( \frac{ e^{(z_i-\mu)/\sigma} }{ 1+e^{(z_i-\mu)/\sigma} } \right)^{Y_i} \left( \frac{ 1 }{ 1+e^{(z_i-\mu)/\sigma} } \right)^{1-Y_i}  \\
  & \times e^{(\mu-\mu_0)^2/(-2\sigma^{2}_{\mu})} e^{-\sigma/\sigma_0} \left( \prod_{j=1}^{p-1} w^{\alpha_j - 1}_j \right) \left( 1-\sum_{j=1}^{p-1} w_j \right)^{\alpha_p -1}
 \end{split}
\end{align}
where $ z_i = w_1 x_{i1} + \cdots + w_{p-1} x_{i,p-1} + w_{p} x_{ip} $, $\sum_{j=1}^{p} w_j=1$ and $ \mathbf{x}_i =(x_{i1}, \cdots ,x_{ip})^{T} $. Then, the MAP is
\begin{align}\label{model-17}{
  \boldsymbol{\hat\theta}_{n} = \arg \max_{\boldsymbol\theta} \log f(\boldsymbol\theta|\mathbf{Y}).
  }
\end{align}
\begin{remark}
Note that a modified Bayesian D-optimal design  in S3 is only used to determine $k_n$ \citep[see][]{DS2008}, and the reason to use it is because of its computational efficiency.
In S4, a more precise criterion $\phi_{1} $ is used to evaluate the candidates found in the previous step.
\end{remark}

\begin{remark}
At early stages, because only few labeled data points are available, the information matrix may be singular, and this is one of the reasons why S4 (ii) is adopted, which is similar to the method used in \citet{DS2008}.
In addition, at the early stage the estimates probabilities of whole unlabeled data points may be close to $ 1 $ or $ 0 $ due to the unstable coefficient estimate.  
It implies that the corresponding uncertainty measure provides little information.  When this situation happens, we will use the distance-based measurement as in \citet{Deng2009} instead until the coefficient estimate becomes stable.
\end{remark}

\color{black}
\section{Simulation Study}
\label{sec: Sim}

In this section, the performance of the proposed method is evaluated through simulation, and compare with that of \citet{Deng2009} with two variables $ \mathbf{x} =(x_{1},x_{2})^{T} $.  (For short, we will refer to their method as ADSL in the rest of this paper.)
We evaluate the performances of two methods with the same misclassification error formulae used in \citet{Deng2009}, which can be estimated by $[\gamma \cdot FP+ (1-\gamma)\cdot FN]/N $, where $ N $ is the total number of data points,  $ FP $ and  $ FN $ are the numbers of the false-positive and false-negative subjects, respectively.
(Note that in \citet{Deng2009},  they only have one parameter, $\alpha$, in their paper.  That is, they have $\omega=\gamma(=\alpha)$ all the time, and let $\alpha=0.5$ when event probability, $P(Y=1)$, is not available.  They also suggested that the parameter $\alpha$ should be adjusted when there is information about the event probability. Note that since they have only one parameter $\alpha$, to adjust $\alpha$ means to adjust both uncertainty measure and cutting threshold.)

We also assess the closeness between the estimated boundaries and the true boundary based on the distance-based measurement used in \citet{Deng2009},  
which is defined as follows:  let
\begin{align}\label{model-19}{
  \mathrm{dist} \equiv \sum_{\mathbf{t}_i \in \mathbf{T}} d_{i}^2,
  }
\end{align}
where $ \mathbf{T} = \{ \mathbf{t}_1,  \mathbf{t}_2, \ldots .\} $ is a set of points that lie evenly on the true boundary, ranging from -3 to 3 on the coordinate of $ x_1 $,  and $ d_i $ is the distance of $ \mathbf{t}_i $ to the estimated boundary for $\mathbf{t}_i \in \mathbf{T}$.
Using (\ref{model-19}), a distance-based performance measure is
\begin{align}\label{model-20}{
  Dist\_PM \equiv \frac{1}{M} \sum_{j=1}^{M} \mathrm{dist}_{j},
  }
\end{align}
where $ M $ is the number of simulations, and $ \mathrm{dist}_{j} $ is the distance defined in (\ref{model-19}) for the $ j $-th simulation.

\subsection{Synthesized Data}
We first compare the propose method with ALSD using a two-dimensional data set with following steps:
\begin{itemize}
\item[](1) Data Generation: We generate simulation data from model (\ref{model-1}) with parameters $ \mu = 0.5 $, $ \sigma = 1 $ and $ w = 0.7 $.  Let $ a_0 = -3 $, $ b_0 = 0 $ and $ \alpha_j = 0.05*j $, where $ j = 1, 2, \cdots, 19 $. We then uniformly  generate $ 5 $ from each interval $ [a_0 + 0.15*(j-1),b_0 + 0.15*(j-1)] $, which are referred to as $ x_1 $. The variable $ x_2 $ is then calculated according to $ F(\mathbf{x}|\boldsymbol\theta) = \alpha_j $. Using $x_1$ and $x_2$, we then generate the response $ Y=1(0) $ with probability $ \alpha_j $ based on the specified logistic model.

\item[](2) Priors: The priors for $ \mu $, $ \sigma $ and $ w $ are described below.
First, consider the prior for $ w $. Assume that the mean of $ w $ to be $ 0.5 $, we set $ \alpha_0/(\alpha_0 + \beta_0 ) = w_0 = 0.5 $, which implies $ \alpha_0 = \beta_0 $. To get a flat prior, we take $  \alpha_0 = \beta_0 =3/2 $.  We then consider the priors for $ \mu $ and $ \sigma $. Based on the lowest and highest value of $ z $ (denoted them as $ z_l $ and $ z_u $) and using formula  $ z = w_0 x_1 + (1-w_0) x_2 $, we choose two extreme points, $ \mathbf{x}_l $ and $ \mathbf{x}_u $. Let $ \alpha_l = 5\% $  and $ \alpha_u = 95\% $ be the suspicious levels for $ \mathbf{x}_l $ and $ \mathbf{x}_u $, respectively. Plugging these values into  (\ref{model-1}),  we have
\[
 \begin{split}
   z_l = \mu + \sigma \log \frac{\alpha_l}{1-\alpha_l},  \hspace{10pt}
   z_u = \mu + \sigma \log \frac{\alpha_u}{1-\alpha_u}.
 \end{split}
\]
Solving the equations above, we obtain $ \mu_0 $ and $ \sigma_0 $ below:
\[
    \mu_0 = \frac{z_l + z_u}{2},  \hspace{10pt}
    \sigma_0 = \frac{z_u - z_l}{\log \frac{\alpha_u}{1-\alpha_u}-\log \frac{\alpha_l}{1-\alpha_l}}.
\]
Take $ \sigma_{\mu}^2 $ as the sample variance of $ z_i $, $ i = 1, \cdots, N $, where $ z_i = w_0 x_{i1} + (1-w_0) x_{i2} $.  Then we complete the prior specification for all three parameters.

\item[](3) For each method, we select points $n = 30 $ sequentially among the total $ 95 $ points. Based on the current labeled points, we estimate the classification function and  calculate the misclassification error and the distance by Equation (\ref{model-19}) for both methods.

\item[](4) Repeat the process $ 100 $ times. The final results are based on the average of 100 runs.
\end{itemize}

\begin{figure}[]
\centering
\includegraphics[width=0.6\textwidth]{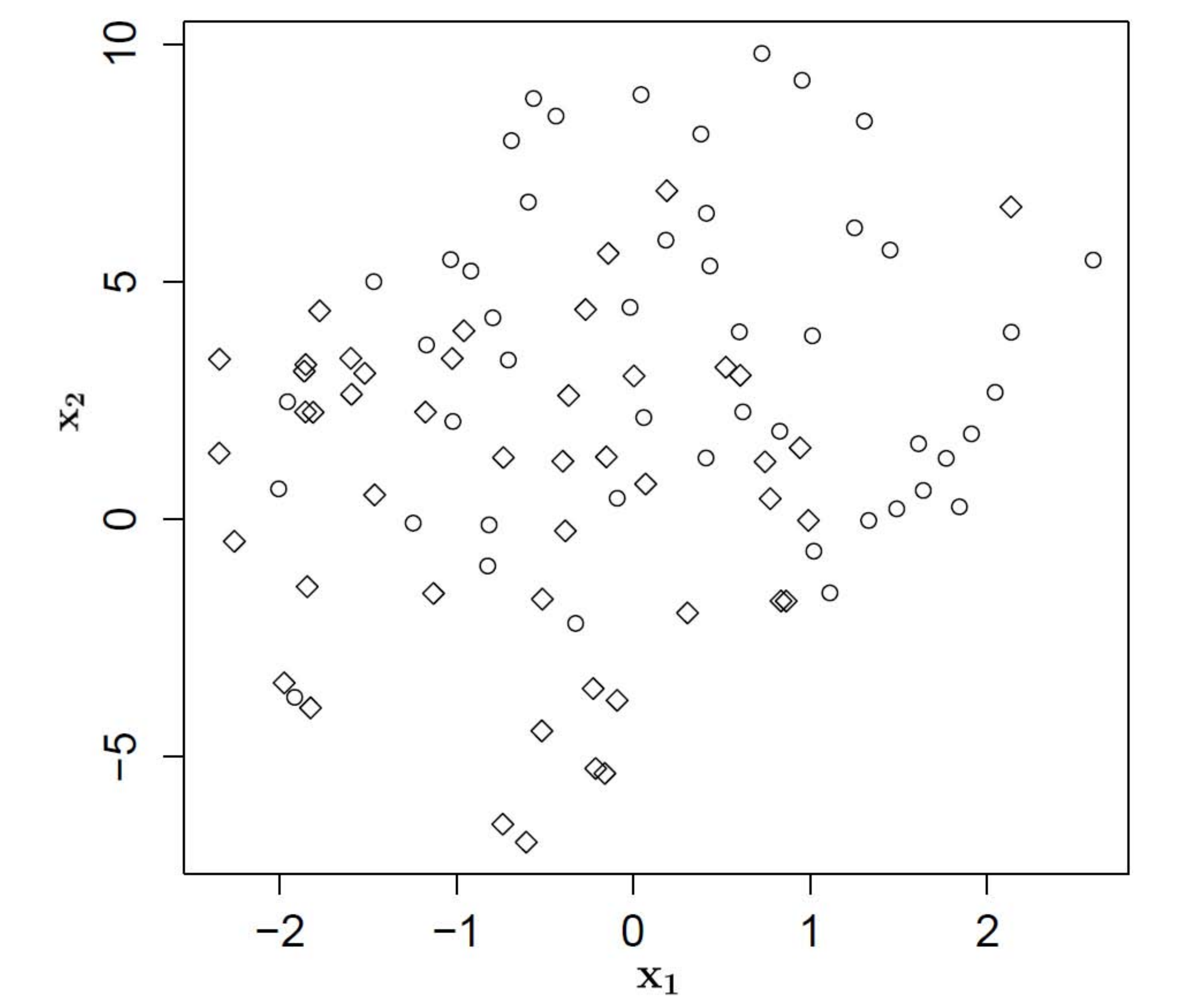}
\caption{Illustration of simulated data; diamond symbol $\diamond$ represents points with response $ Y=0 $ and circle symbol $\circ$ denotes points with response $ Y=1 $.}
%\todo[inline]{this figure is not clear enough.  Re-draw it using  larger symbols.  It should be save as pdf file in the first place.  Not transfer from jpeg to pdf.  The resolution is too low.}
\label{Fig:toy-data}
\end{figure}

According to the previous design,  an example set of $ N = 95 $ simulated data points is illustrated in Figure \ref{Fig:toy-data}, where circle and square denote two different groups.
From this figure,  we can see that the two labeled points are mixed together, and the range of $ x_1 $ remains in $ (-3,3) $. The responses with $ Y = 1 $ are observed mostly when two explanatory variables are both large.

\color{black}
\subsubsection{Results}

\begin{figure}[]
          \centering
            \subfloat[]{
	            \begin{minipage}[t]{0.5\linewidth}
	            \includegraphics[width=7.5cm,height=7.5cm]{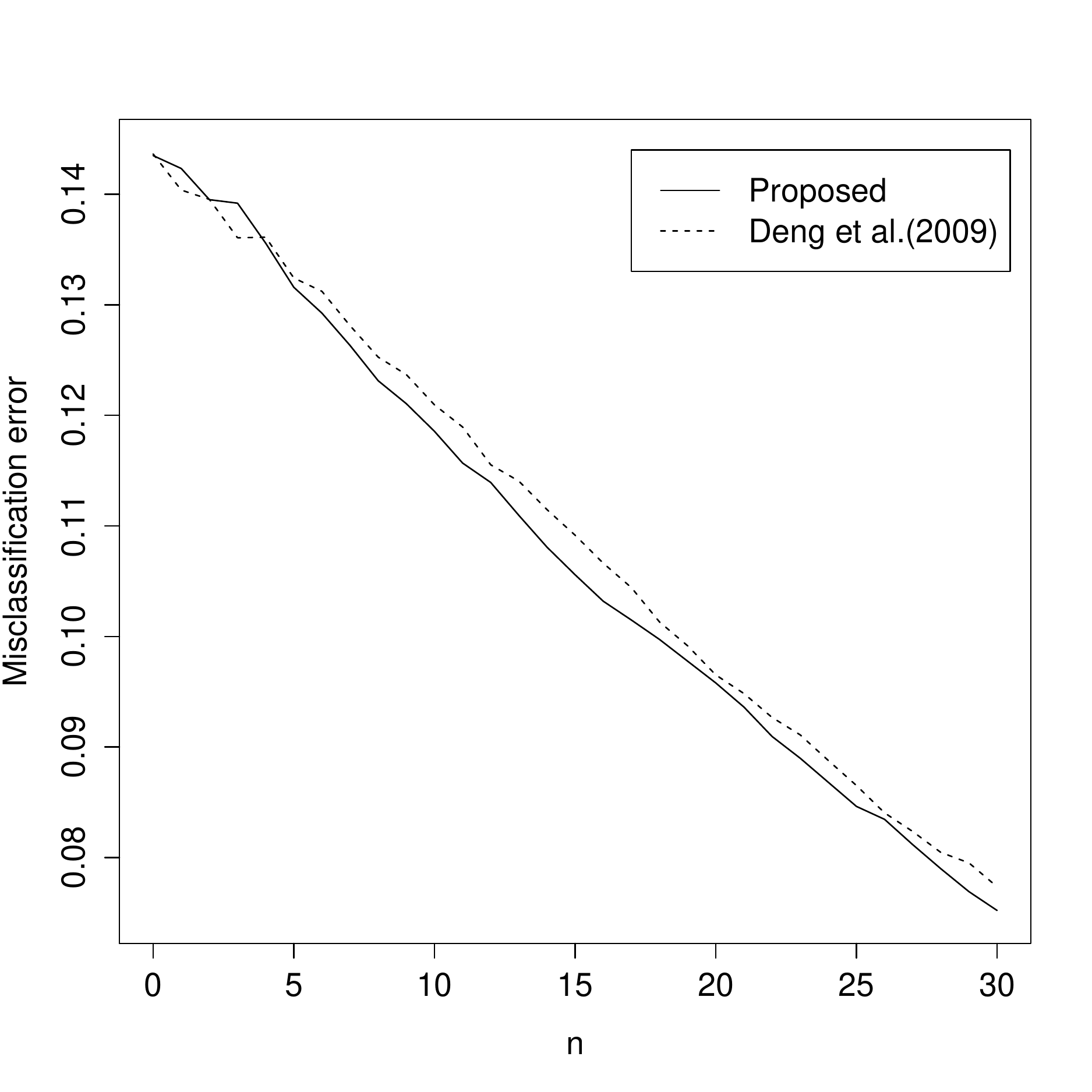}
	            \centering
	            %\caption{(a)}
	            %\label{fig:notification-sys:a}
	            \end{minipage}
	        }
            \subfloat[]{
	            \begin{minipage}[t]{0.5\linewidth}
	            \centering
	            \includegraphics[width=7.5cm,height=7.5cm]{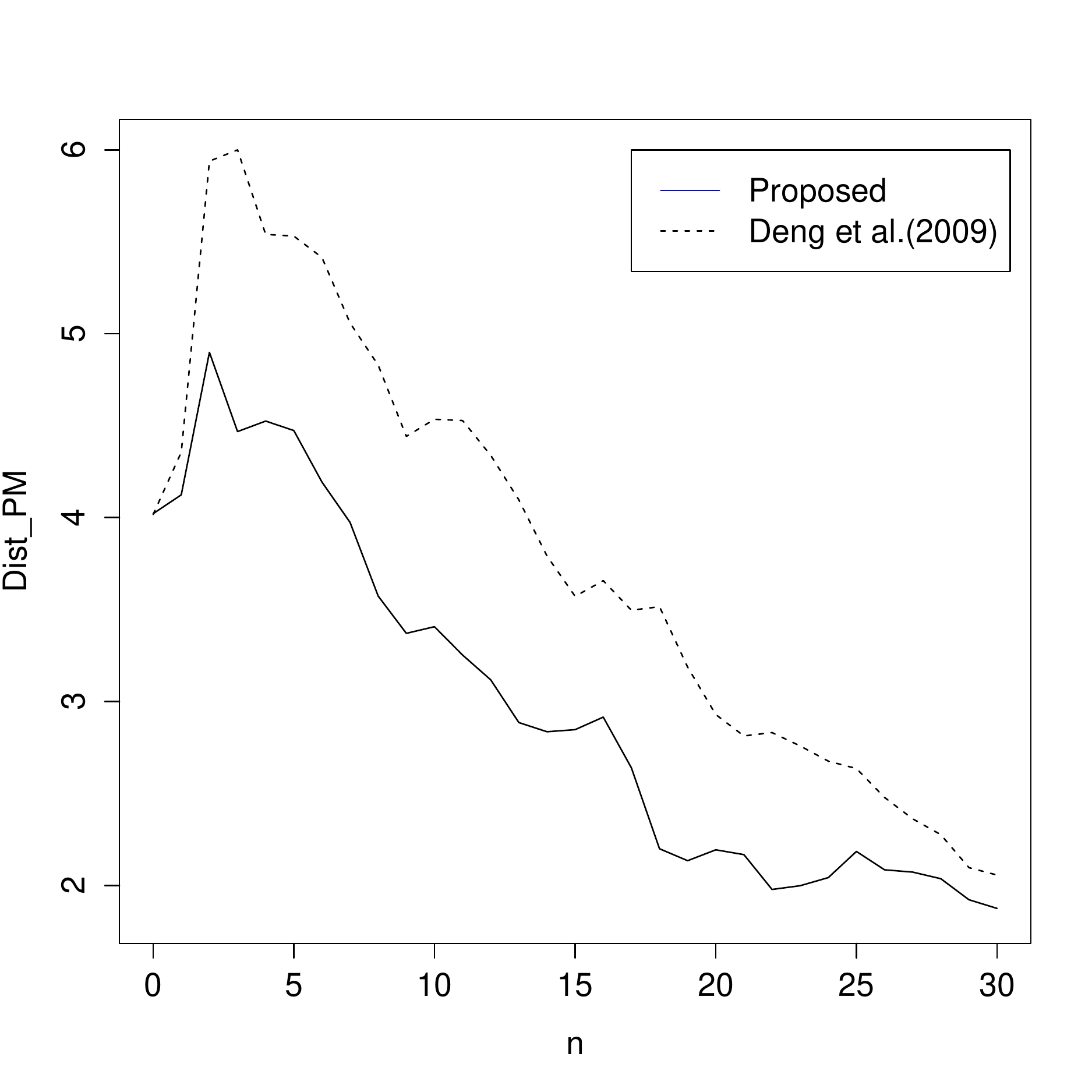}
	            %\caption{(b)}
	           % \label{fig:notification-sys:b}
	            \end{minipage}
            }
          \caption{Learning curves of the proposed method compared with \citet{Deng2009}. (a) Misclassification Error; (b) Distance-based Measure.}
          \label{Fig:sim}
\end{figure}

Based on 100 runs, curves of the misclassification error and distance-based measure are shown in Figure \ref{Fig:sim} (a) and (b), respectively. The misclassification errors of the proposed method are slightly smaller than those of ALSD starting from around $ n > 4 $. The estimated boundary of the proposed method also moves towards the true boundary faster than that of ALSD.  Note that the number of candidate set in ALSD is $ k_0 = 20 $, and fixed for all stages.  Because $p=3$ in our simulation, the number of candidate set for the proposed method is $ k_n < 4p ( = 12) $, which vary according to the criterion of the local efficiency method mentioned before and is smaller than $ k_0 $.  Hence, we only have to access less candidate points and is usually computational more efficient.

%\vspace{-0.5cm}
\color{black}
\subsection{Advantages of a small amount of initial learning subjects}
\begin{figure}[]
    %      \centering
            \subfloat[ $ n_0 = 5 $ ]{
	            \begin{minipage}[t]{ 0.5\linewidth }
	            \includegraphics[width=7cm,height=5cm]{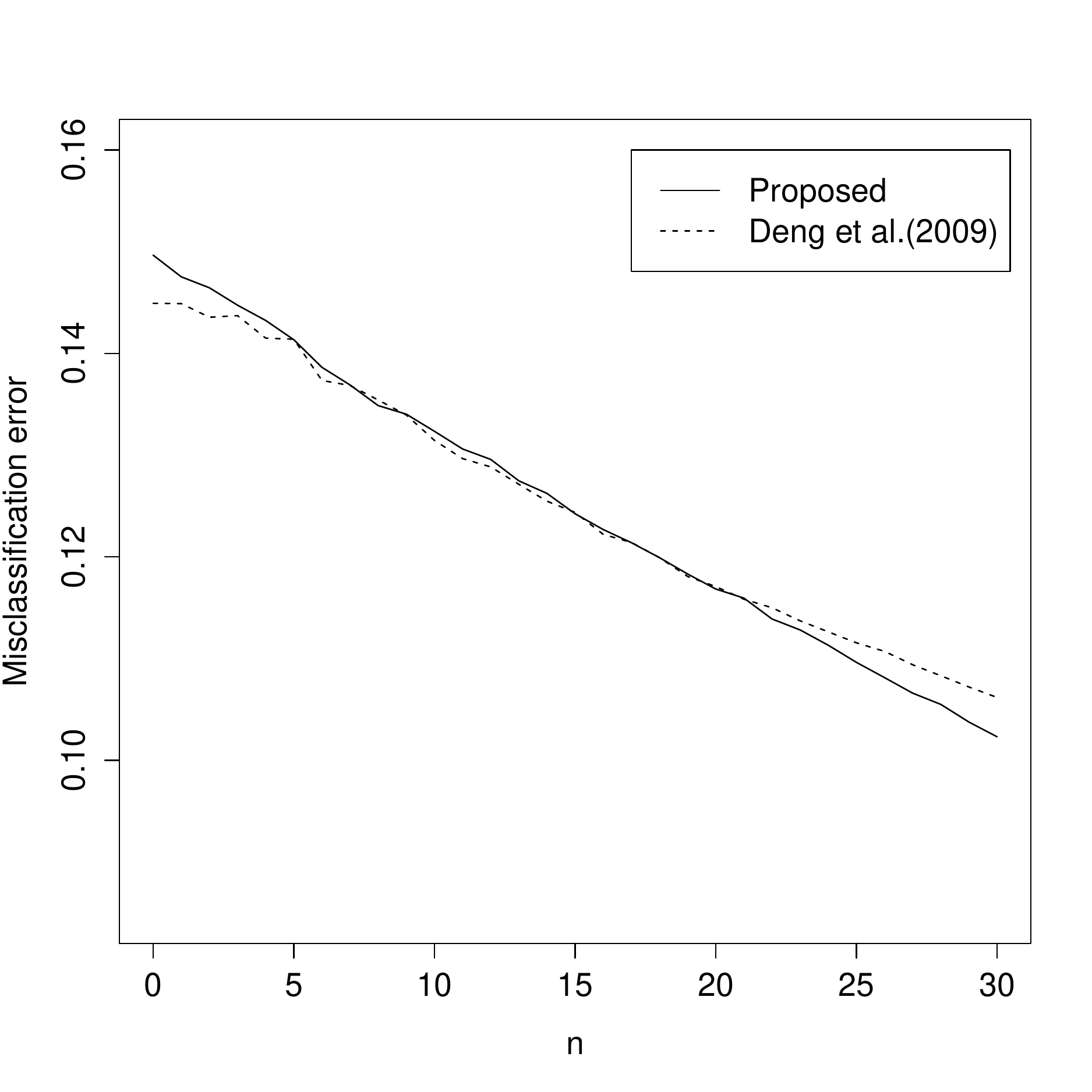}
	            \centering
	            %\caption{(a)}
	            %\label{fig:notification-sys:a}
	            \end{minipage}
	        }
            \subfloat[ $ n_0 = 10 $ ]{
	            \begin{minipage}[t]{ 0.5\linewidth }
	            \centering
	            \includegraphics[width=7cm,height=5cm]{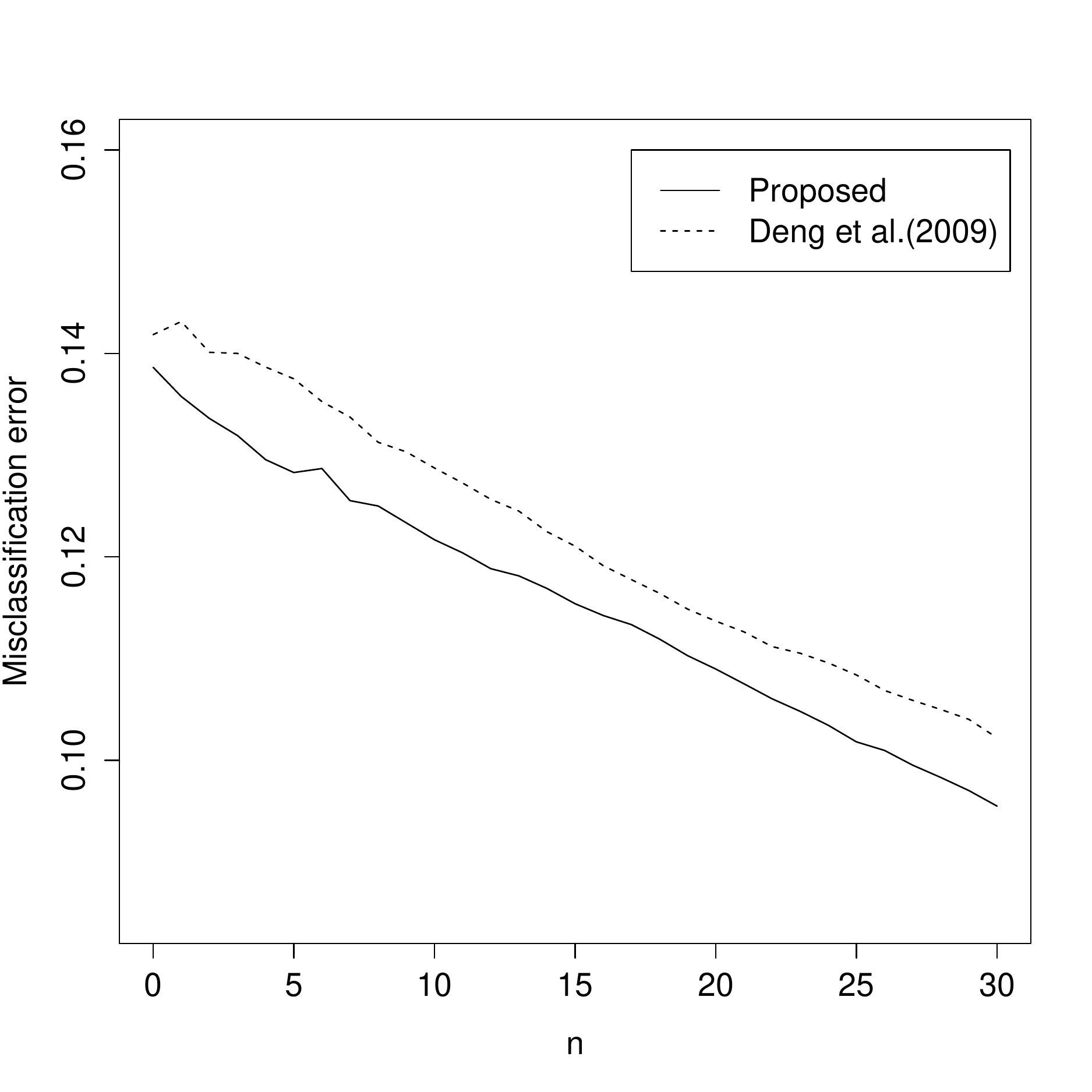}
	            %\caption{(b)}
	           % \label{fig:notification-sys:b}
	            \end{minipage}
            }  \\
            \subfloat[ $ n_0 = 15 $ ]{
	            \begin{minipage}[t]{ 0.5\linewidth }
	            \includegraphics[width=7cm,height=5cm]{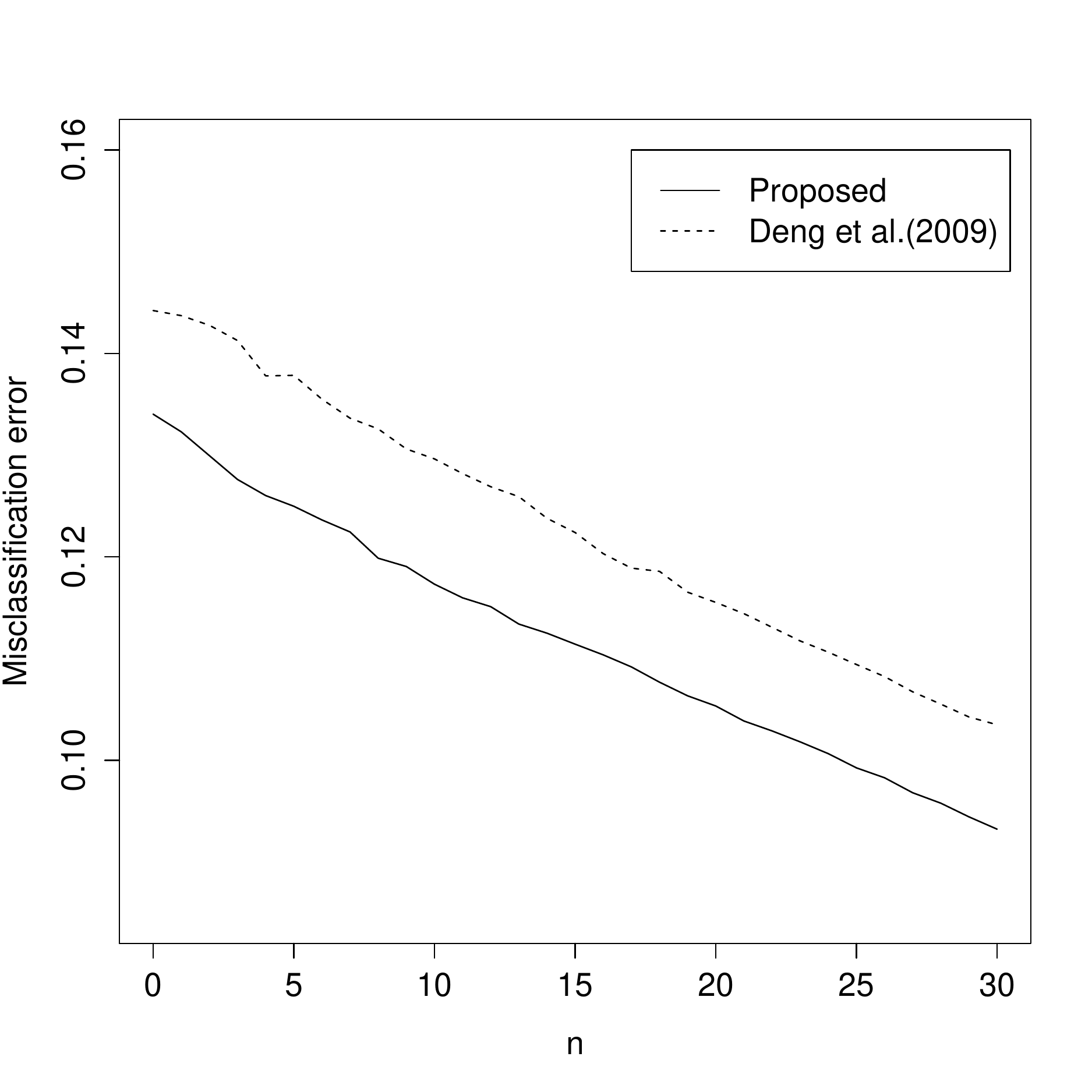}
	            \centering
	            %\caption{(a)}
	            %\label{fig:notification-sys:a}
	            \end{minipage}
	        }
            \subfloat[ $ n_0 = 20 $ ]{
	            \begin{minipage}[t]{ 0.5\linewidth }
	            \centering
	            \includegraphics[width=7cm,height=5cm]{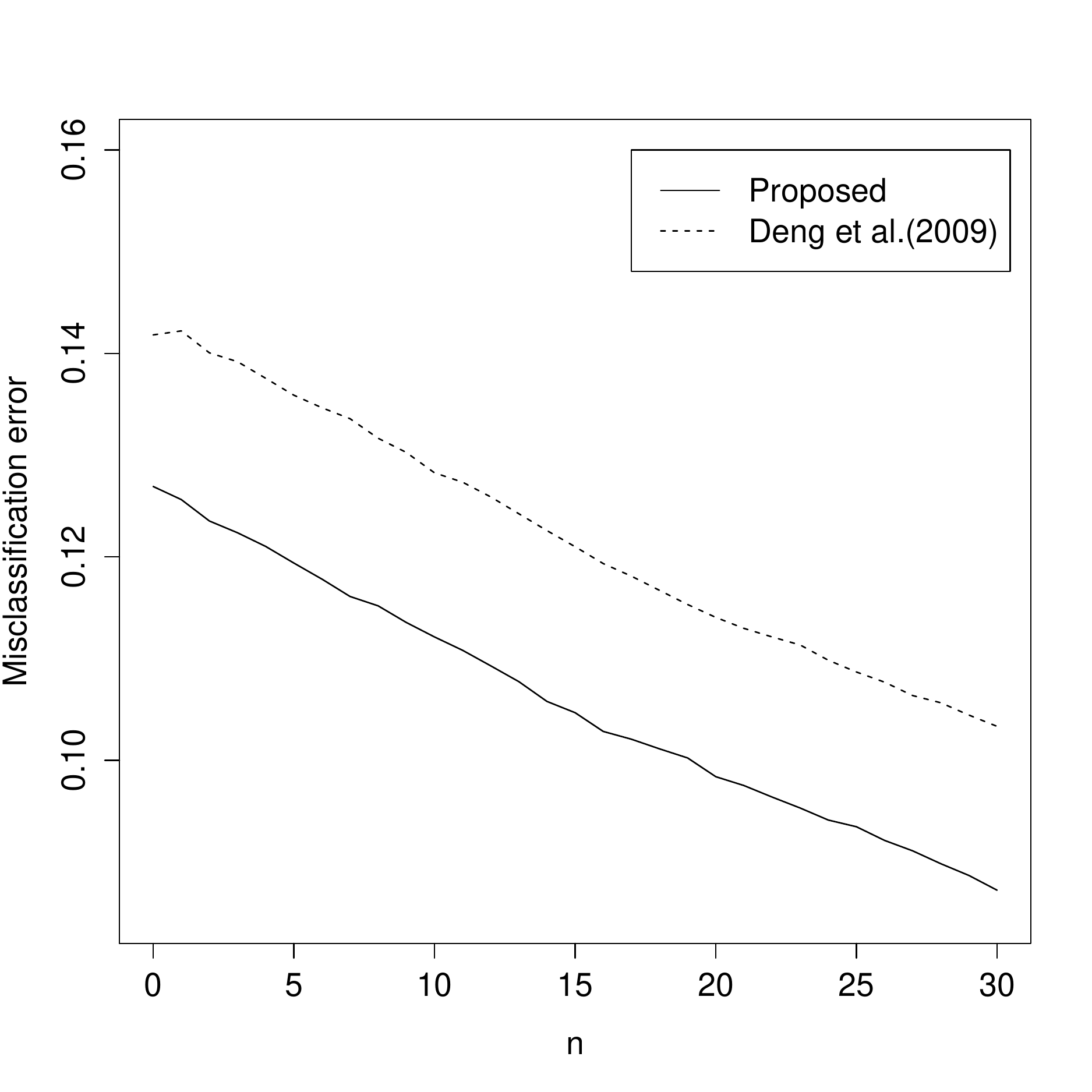}
	            %\caption{(b)}
	           % \label{fig:notification-sys:b}
	            \end{minipage}
            }
          \caption{Comparing the misclassification error curves of the proposed method with different sizes of the labeled data points as its initial training set ($n_0=5, 10, 15, 20$) to that of the method of \citet{Deng2009} with $n_0=0$.  }
          \label{Fig:sim-n0}
\end{figure}

\citet{Deng2009} did not discuss the effects of using more than one labeled subject as an initial training set.
However, because active learning algorithms are sequential procedures, the performance of the current stage relies on the information obtained from its predecessors. 
Hence, how to have a good and stable early performance in stages will play an important role in a successful active learning process.  
An easy way to have a good start is to have more labeled samples in its initial stage.  
Thus, to see the effects of different initial training sizes, we generate a data set with $ N = 190 $ data points, and compare the results of the proposed method with $ n_0 $ equal to $ 5, 10, 15, 20$  to that of ALSD with $n_0=0$.  
Figure \ref{Fig:sim-n0} shows misclassification curves of both the proposed methods with $n_0 = 5, 10, 15, 20$, respectively, and ALSD with $n_0=0$. 
As expected the proposed method with $n_0 > 0$ performs better than ALSD, and  the larger the initial training size $ n_0 $, the better performance of the proposed method.  
In  Figure \ref{Fig:sim-n0} (d), for example, shows that at around $n=15$ labeled samples, 
the proposed method can achieve the same classification performance of ALSD  at $n_0= 30$. 
Because the computation of active learning is time consuming process at each stage, and this is especially the case in problems with high dimensional data.
Thus, a method requires less learning stages can help to save the computational time.  
Hence,  in order to ensure a stable and efficient learning process, it is recommended to start with a small amount of labeled data, if they can be available.   
In fact, for some cases, we actually require less total number of the labeled subjects to achieve the same performance of ALSD.
This situation can be seen from some real data examples and will be discussed later.

\color{black}
\section{Real Examples}
\label{sec: Real}

For illustration and comparison purposes, we apply both the proposed method and ALSD to Liver Disorders (BUPA) and Wisconsin Diagnostic Breast Cancer (WDBC) data sets, which are available at the UCI Repository of Machine Learning Databases \citep{Bache+Lichman:2013}.  Our main interest is correct classification rate, so we use the same misclassification error formulae defined before to evaluate their performances.

\subsection{BUPA data set}

\begin{figure}[]
    %      \centering
            \subfloat[ $ n_0 = 0 $ ]{
	            \begin{minipage}[t]{ 0.5\linewidth }
	            \includegraphics[width=7cm,height=5cm]{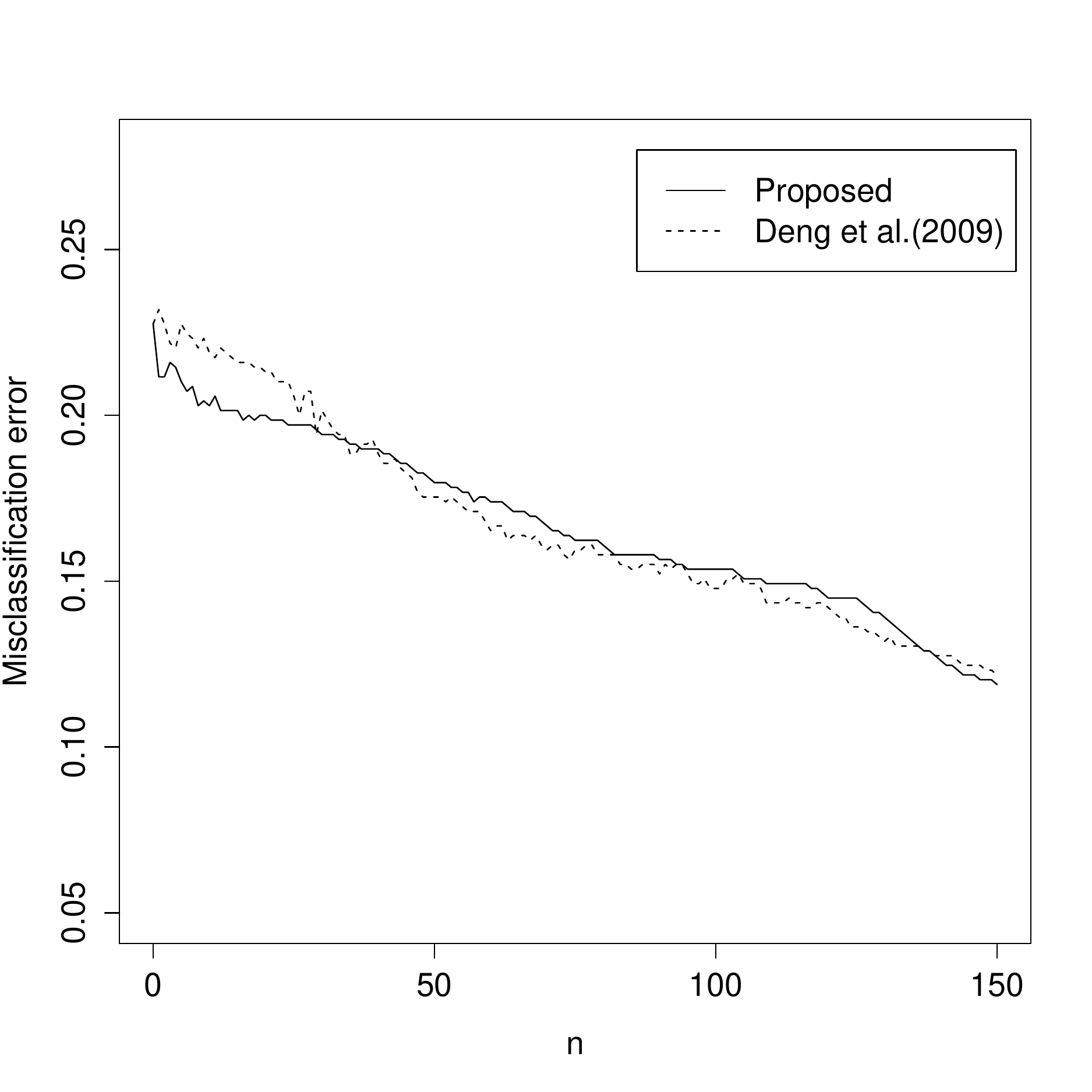}
	            \centering
	            %\caption{(a)}
	            %\label{fig:notification-sys:a}
	            \end{minipage}
	        }
            \subfloat[ $ n_0 = 10 $ ]{
	            \begin{minipage}[t]{ 0.5\linewidth }
	            \centering
	            \includegraphics[width=7cm,height=5cm]{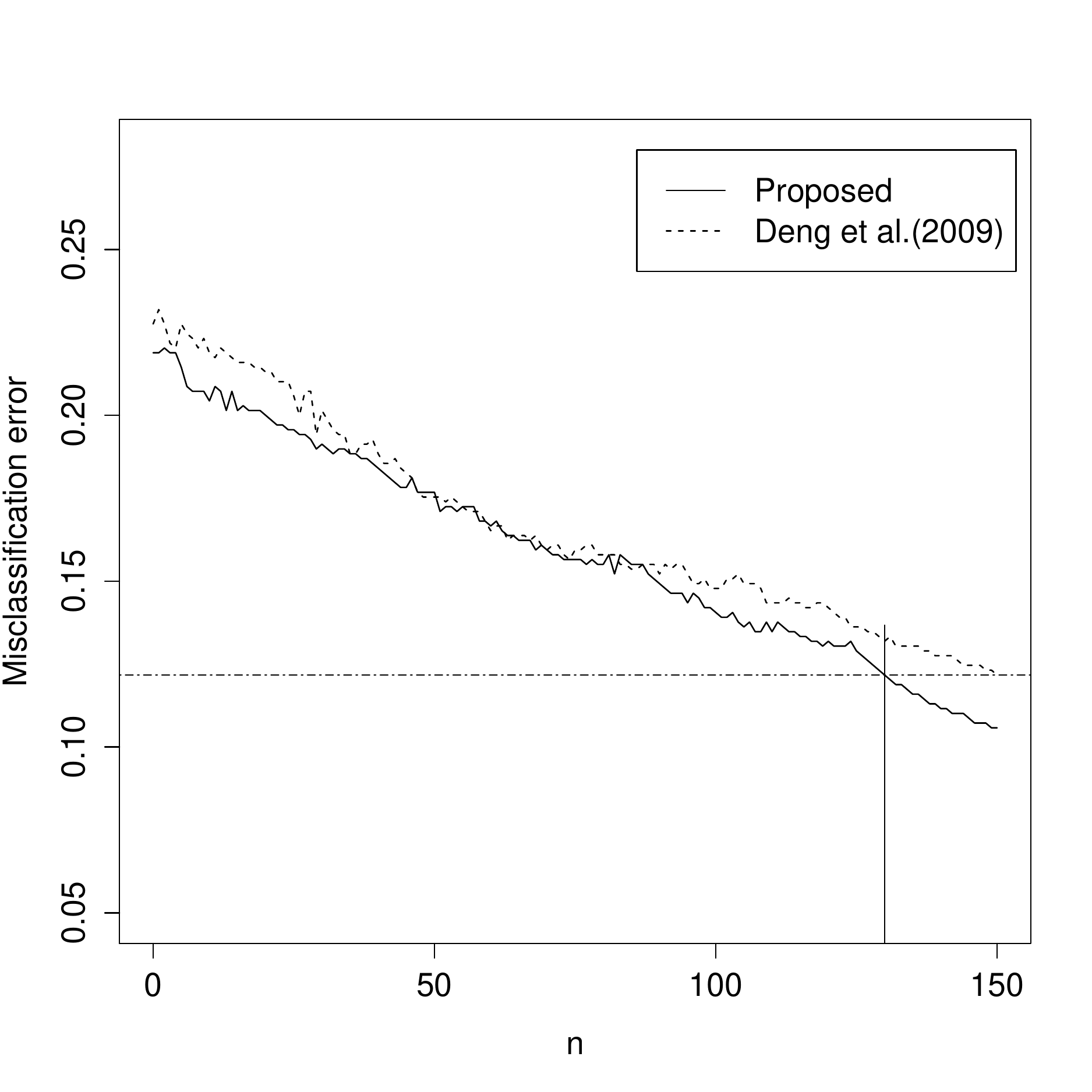}
	            %\caption{(b)}
	           % \label{fig:notification-sys:b}
	            \end{minipage}
            }  \\
            \subfloat[ $ n_0 = 20 $ ]{
	            \begin{minipage}[t]{ 0.5\linewidth }
	            \includegraphics[width=7cm,height=5cm]{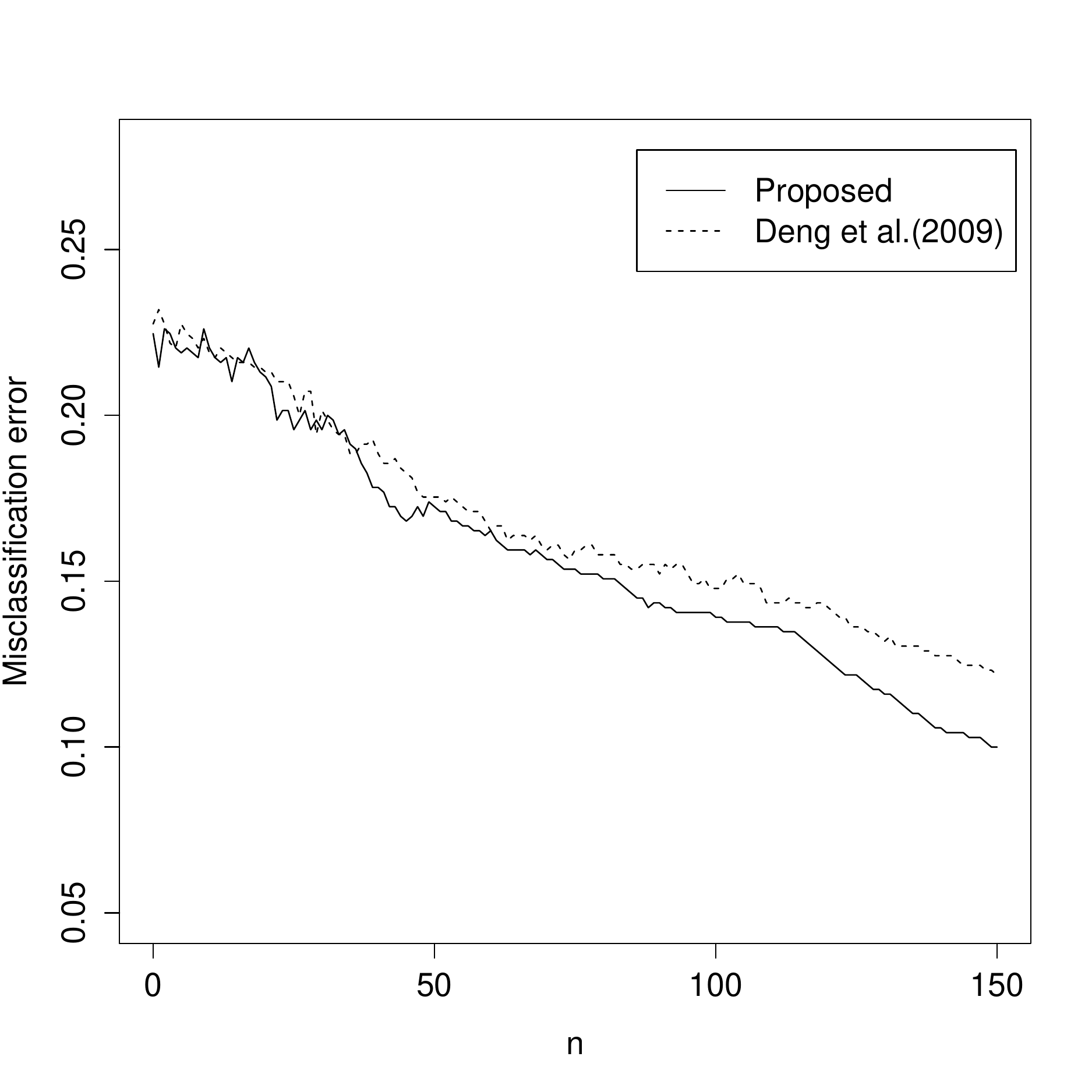}
	            \centering
	            %\caption{(a)}
	            %\label{fig:notification-sys:a}
	            \end{minipage}
	        }
            \subfloat[ $ n_0 = 30 $ ]{
	            \begin{minipage}[t]{ 0.5\linewidth }
	            \centering
	            \includegraphics[width=7cm,height=5cm]{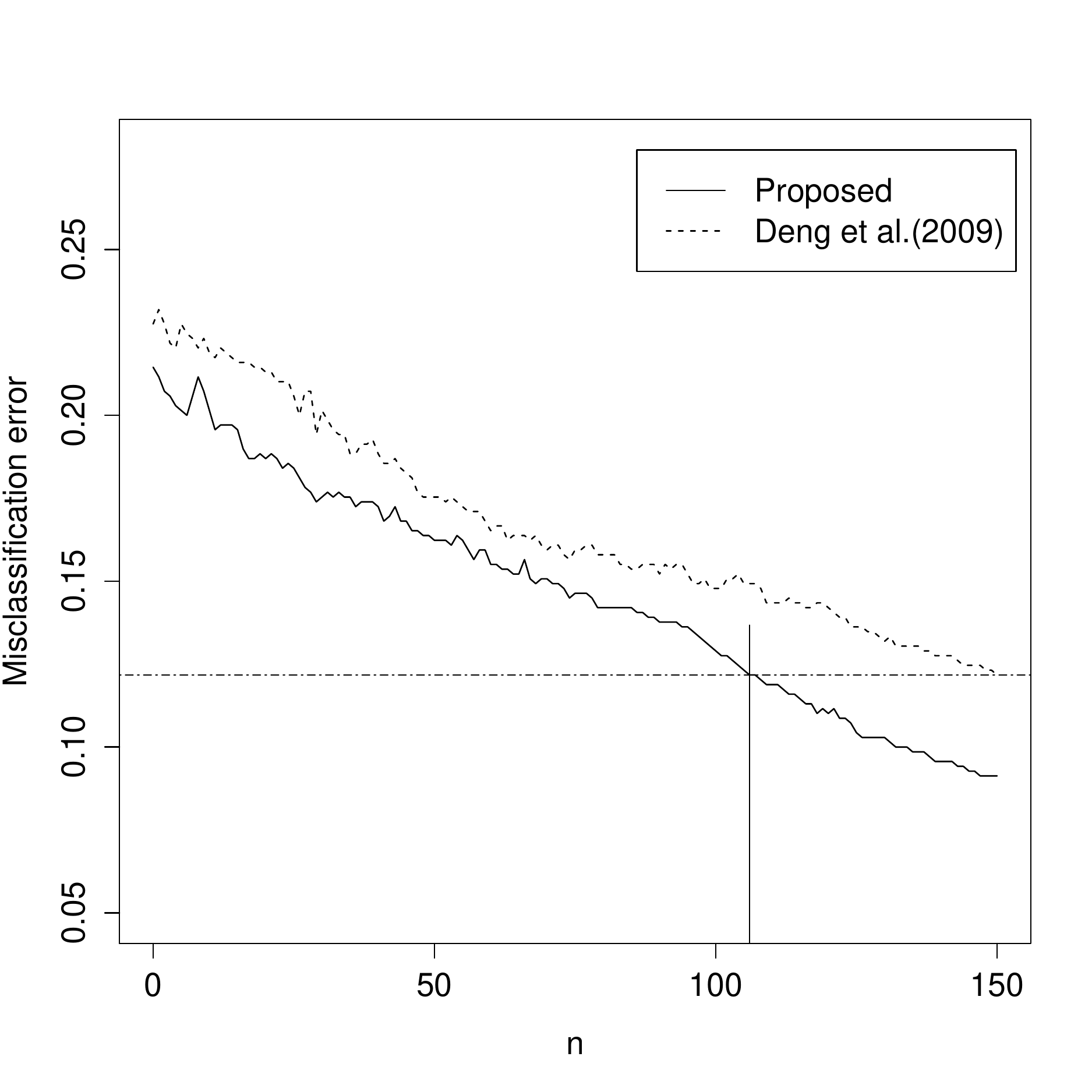}
	            %\caption{(b)}
	           % \label{fig:notification-sys:b}
	            \end{minipage}
            }
          \caption{Misclassification Error curves for the two methods, using BUPA data set with $ n_0 =0,  10, 20, $ and $ 30 $.}
          \label{Fig:BUPA-n0}
\end{figure}

The original BUPA data set is from the California state, USA, which contains 345 records (145 liver patient and 200 non-liver patient records) with 6 attributes as shown in Table \ref{tab:BUPA}.  The first 5 variables are from blood tests and sensitive to liver disorders, which  might be due to excessive alcohol consumption. 
All features are positive related to the response in a general sense. 
That is, the higher the value of variables, the higher the probability that the corresponding subject is liver disordered. 
The performances of the two methods (the proposed one and ALSD) in terms of misclassification error are illustrated in Figure \ref{Fig:BUPA-n0}. 
Our  method performs similarly to ALSD when $n_0=0$.  
However, in the proposed method, because $p=7$, we only need to evaluate $k_n (\leq 4p=28)$ candidates at each stage  which is smaller than the number of candidates ($k_0=50$) used  in ALSD.  That is,  we only have to access a smaller number of candidates, which will save us a lot of computational time.

\begin{table}[!htbp]
\centering
   \begin{tabular}{lll}
   \hline
   Attribute       &      Type      &      Detail                                   \\
   \hline
   Mcv             &     Integer    &    Mean corpuscular volume                    \\

   Alkphos         &     Integer    &    Alkaline phosphotase                       \\

   SGPT            &     Integer    &    Alamine aminotransferase                   \\

   SGOT            &     Integer    &    Aspartate aminotransferase                 \\

   Gammagt         &   Real   &    gamma-glutamyl transpeptidase              \\

   Drinks          &   Real   &    number of half-pint equivalents of         \\
                   &                &    alcoholic beverages drunk per day          \\
   \hline
   \end{tabular}
\caption{Attribute information for BUPA data set.}
\label{tab:BUPA}
\end{table}

As in the previous section, we also start with different $ n_0 (\geq 0)$ as an initial data set.  The total size of BUPA is $ N = 345 $ (about $ 2\times190 $),  and we set $n_0 = 0, 10, 20$, and $ 30$. Figure \ref{Fig:BUPA-n0} shows that our method performs better than  the ALSD as $n_0$ gets larger, and the difference of two curves increases as $n_0$ increases. It is worth to note that even with $n_0=10$ at around 130 training samples, the proposed method can achieve the same classification performance of ALSD at 150 labeled data points.  That is, it saves about 10 labeled samples in total in this case.  Similar situations can be found in the cases with other $n_0$'s. For example, with $n_0=30$, the proposed method requires only, in average,  110 labeled subjects to achieve the performance of ADSL with 150 labeled samples.  Because it is a sequential process, it implies that the proposed method requires less training stages to achieve the same performance level, and therefore is more efficient in terms of training time.
In practice,  there will be some cost for experts to label subjects. Thus, to save labeled samples is not only to save learning time, but also the budget of a learning process.

\subsection{Application to WDBC data set}
The WDBC data set contains $569$ breast masses with $ 357 $ benign and $ 212 $ malignant cases. Ten different features are measured including radius, perimeter, area, compactness, smoothness, concavity, concave points, symmetry, fractal dimension and texture. All features are numerically modeled such that larger values are typically indicated a higher likelihood of malignancy \citep[see][]{street1993nuclear}. The details can  also be found in \citet{wolberg1994machine}, and \citet{mu2008breast}.
The mean value, extreme (largest or ``worst'') value and standard error of each feature are computed for each image, which resulted in a total of 30 features of 569 images, and yielded a database of $ 569 \times 30 $ samples.

\begin{figure}[]
    %      \centering
            \subfloat[ $ n_0 = 0 $ ]{
	            \begin{minipage}[t]{ 0.5\linewidth }
	            \includegraphics[width=7cm,height=5cm]{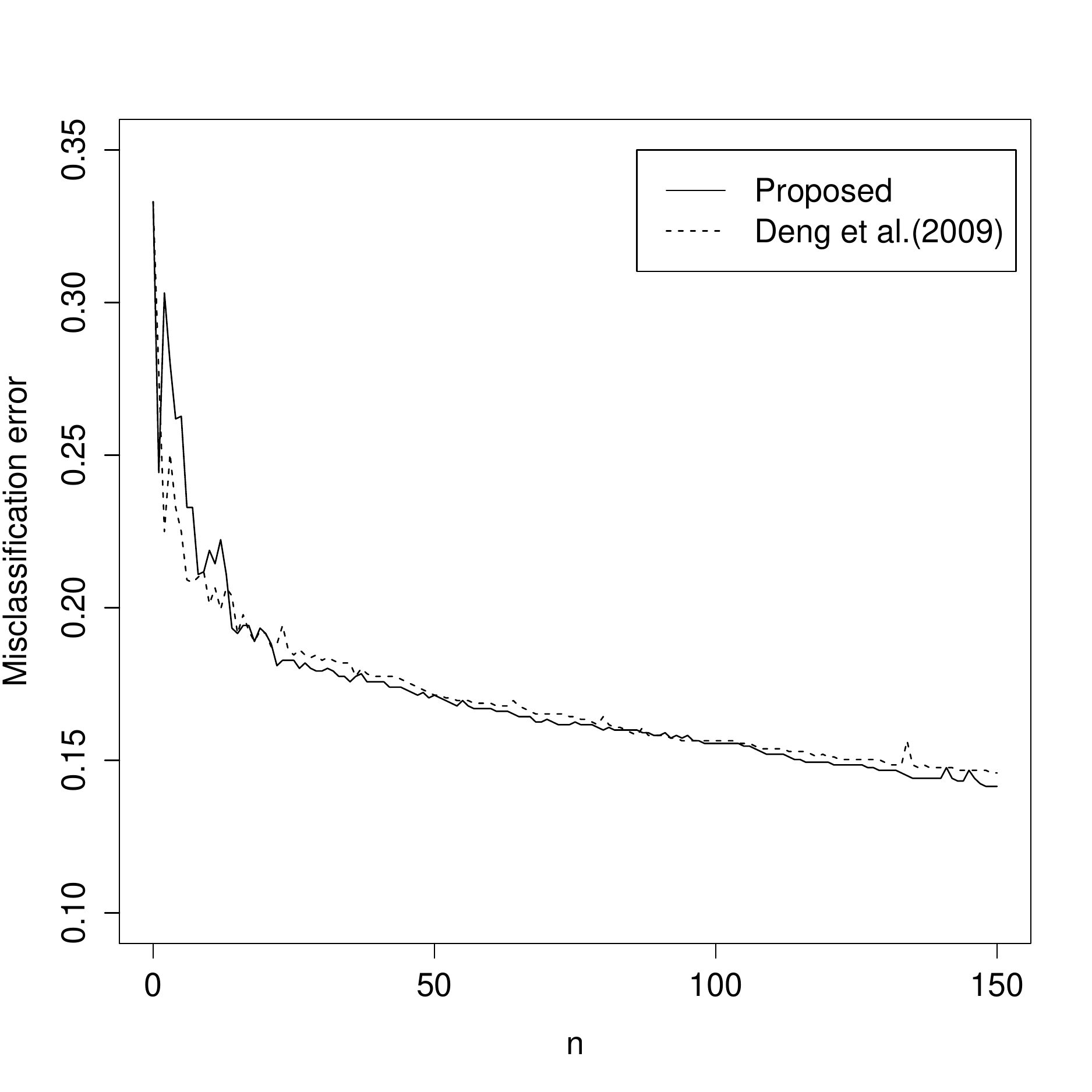}
	            \centering
	            %\caption{(a)}
	            %\label{fig:notification-sys:a}
	            \end{minipage}
	        }
            \subfloat[ $ n_0 = 15 $ ]{
	            \begin{minipage}[t]{ 0.5\linewidth }
	            \centering
	            \includegraphics[width=7cm,height=5cm]{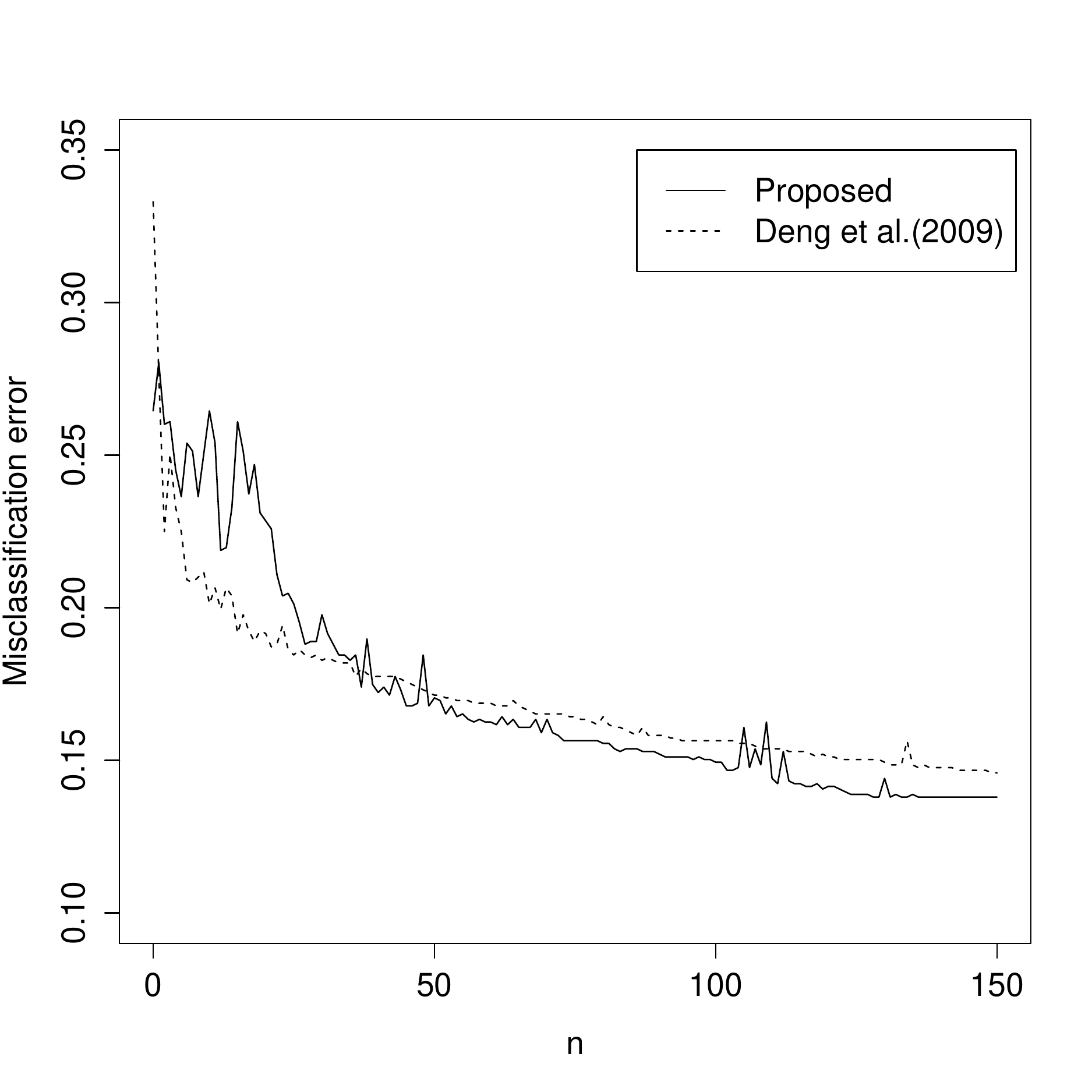}
	            %\caption{(b)}
	           % \label{fig:notification-sys:b}
	            \end{minipage}
            }  \\
            \subfloat[ $ n_0 = 30 $ ]{
	            \begin{minipage}[t]{ 0.5\linewidth }
	            \includegraphics[width=7cm,height=5cm]{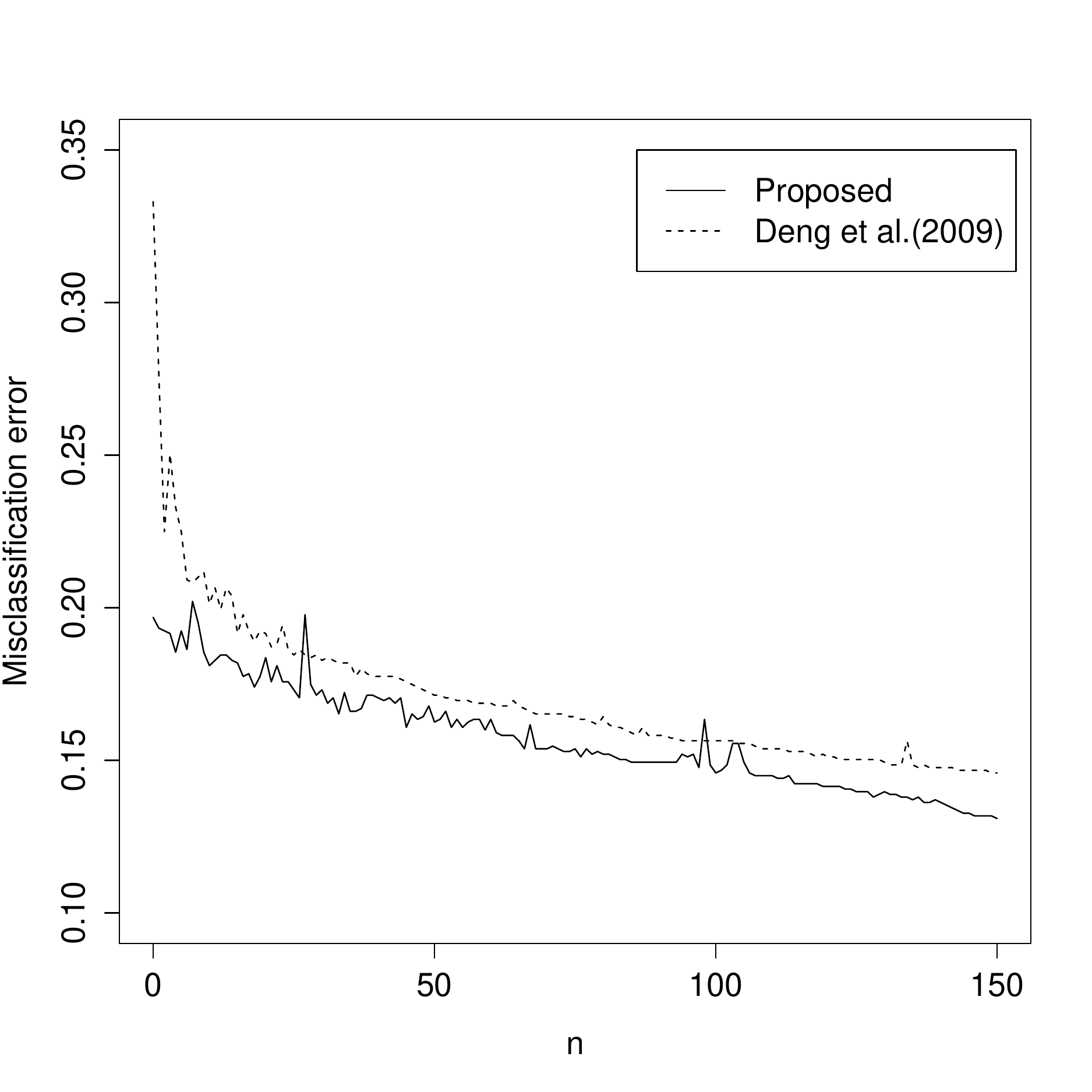}
	            \centering
	            %\caption{(a)}
	            %\label{fig:notification-sys:a}
	            \end{minipage}
	        }
            \subfloat[ $ n_0 = 45 $ ]{
	            \begin{minipage}[t]{ 0.5\linewidth }
	            \centering
	            \includegraphics[width=7cm,height=5cm]{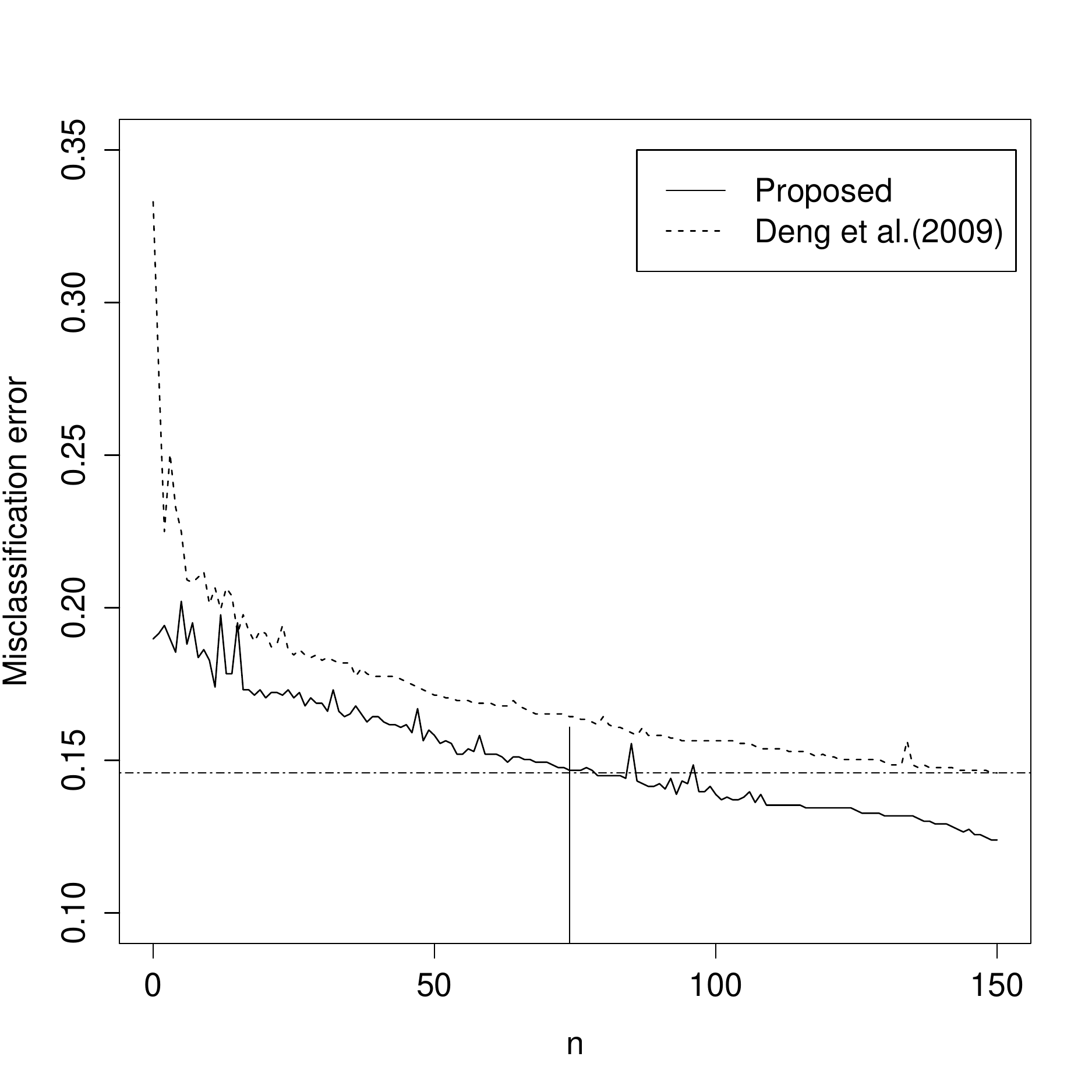}
	            %\caption{(b)}
	           % \label{fig:notification-sys:b}
	            \end{minipage}
            }
          \caption{Misclassification Error curves for the two methods using WDBC  with $ n_0 = 0, 15, 30$ and  $ 45 $.}
          \label{Fig:WDBC-n0}
\end{figure}

We apply both the proposed method and ALSD to WDBC data set, and their misclassification error curves are shown in Figure \ref{Fig:WDBC-n0} with different $n_0=0, 15, 30$ and $45$ for the proposed method and $n_0=0$ for ALSD.
When the initial training sample size $n_0$ increases, the proposed method outperforms ALSD as expected.  
It is worth to note that the misclassification errors for processes starting with a small amount of labeled data ($n_0=30, 45$) are smaller than that of ALSD from the very beginning.  It also shows that with a small amount of initial training subjects the proposed method achieves the same classification performance  sooner than ALSD at a stage with less labeled samples.  For example, in \ref{Fig:WDBC-n0} (d), with around 75 to 90 labeled samples, the proposed method achieves a misclassification error that is similar to that of ALSD with 150 labeled  samples.  Hence, even using 45 labeled subjects at the initial stage, we still save about 15 to 30 subjects.  Thus, to start with a small amount of labeled samples as an initial training set will actually be more efficient in both cost and computational time.

\subsection{Active learning when group sizes are uneven}
\label{sec:uneven}
\begin{figure}[]
          \centering
            \subfloat[]{
	            \begin{minipage}[t]{0.5\linewidth}
	            \includegraphics[width=7cm,height=5cm]{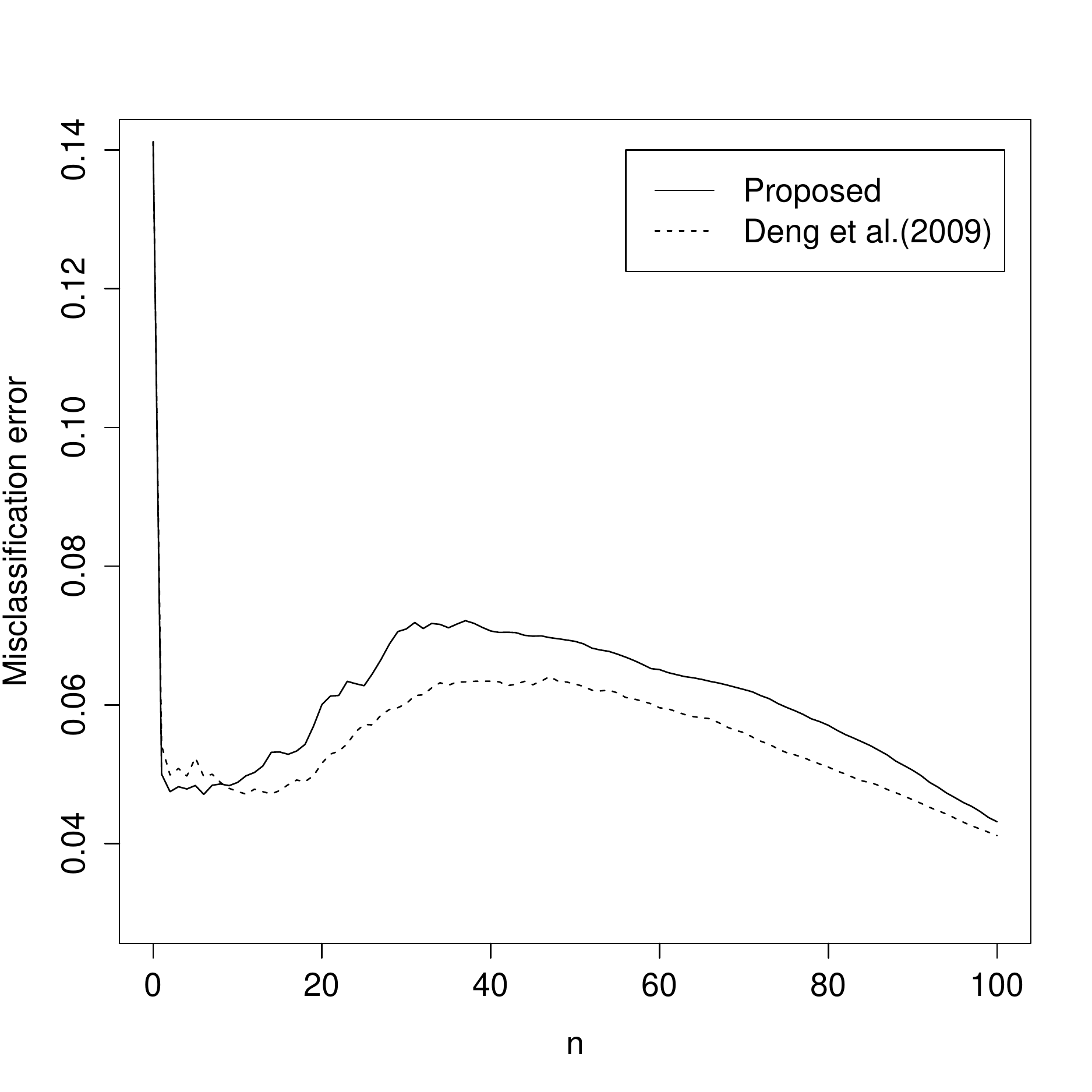}
	            \centering
	            %\caption{(a)}
	            %\label{fig:notification-sys:a}
	            \end{minipage}
	        }
            \subfloat[]{
	            \begin{minipage}[t]{0.5\linewidth}
	            \centering
	            \includegraphics[width=7cm,height=5cm]{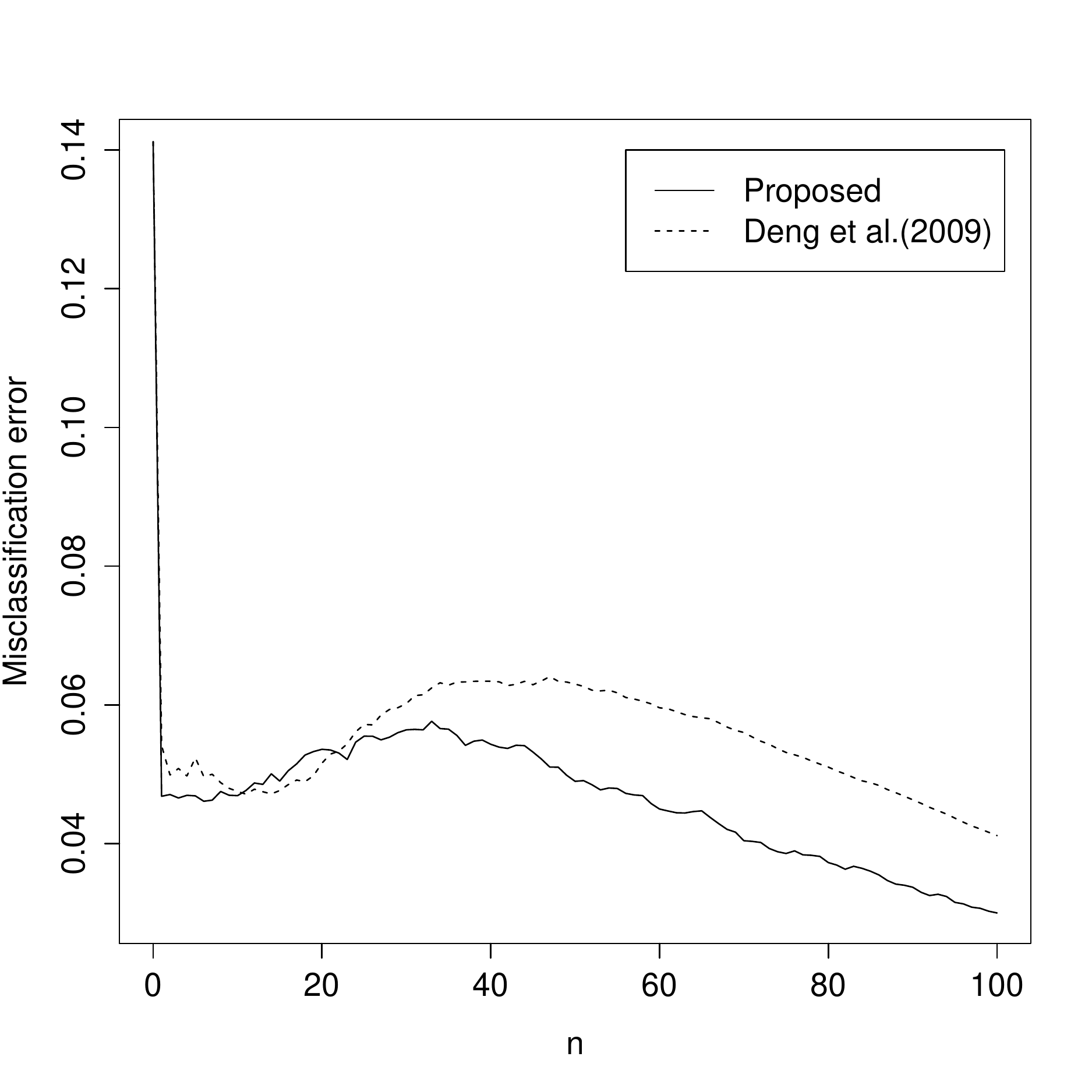}
	            %\caption{(b)}
	           % \label{fig:notification-sys:b}
	            \end{minipage}
            }
          \caption{Comparing the misclassification error curves of the proposed method with \citet{Deng2009} when sample sizes of two groups are uneven with different choices of uncertainty level $\omega$s and sample size dependent cutting point $\gamma$s.
          All methods start from $n_0=0$.  (a) Misclassification curves of two methods with both uncertainty and cutting point are equal to 0.8;  (b) Here, the proposed method sets uncertainty probability $\omega = 0.5$ and a cutting point $\gamma= 0.8$, while ALSD uses $\omega=\gamma=0.5$. Note that the misclassification of the proposed method $(< 0.04)$ in (b) is also smaller than that of ALSD $(> 0.04)$ in (a) when $n=100$. }
          \label{Fig:uneven}
\end{figure}

\paragraph{Synthesized Data}
When  either the ratio of two group size or the odds ratio of two groups is extreme, then the classification rule should take this information into consideration.  \citet{Deng2009} suggested using $F(x)=\omega$ and adjusted $\omega$ based on the probability of a case if there is a prior information available. (Note that in \citet{Deng2009}, they used the same number, denoted as $\alpha$, in both uncertainty measurement and event probability adjustment.)
In this section, we conduct some numerical studies with uneven group sizes. The results in Figure \ref{Fig:uneven} are based on simulated data with size ratio equals to 1 to 4.
We first set both the uncertainty probability ($\omega$) and cutting point ($\gamma$) of the proposed method equal to 0.8, and the $\alpha=0.8$ in ALSD. It can be seen from Figure \ref{Fig:uneven}  (a) , that the misclassification rate of the proposed method under such a setup is worse than that of ALSD with $\alpha=0.8$.  However, if we set the uncertainty sampling probability equal to $\omega=0.5$ and  adjust the uneven group sizes with a shift cutting point based on the ratio of two sample sizes (i.e. $\gamma=0.8$), then the performance of the proposed method is much improved (see Figure \ref{Fig:uneven}  (b)), and is better than that of ALSD.  (All the learning curves in Figure \ref{Fig:uneven}  are based on the average of 100 replications of each method.)   We also conducted simulations for other group size ratios, and the results are all similar and therefore omitted here.

\begin{figure}[]
            \subfloat[ BUPA data with $ n_0 = 0 $ ]{
	            \begin{minipage}[t]{ 0.5\linewidth }
	            \includegraphics[width=7cm,height=5cm]{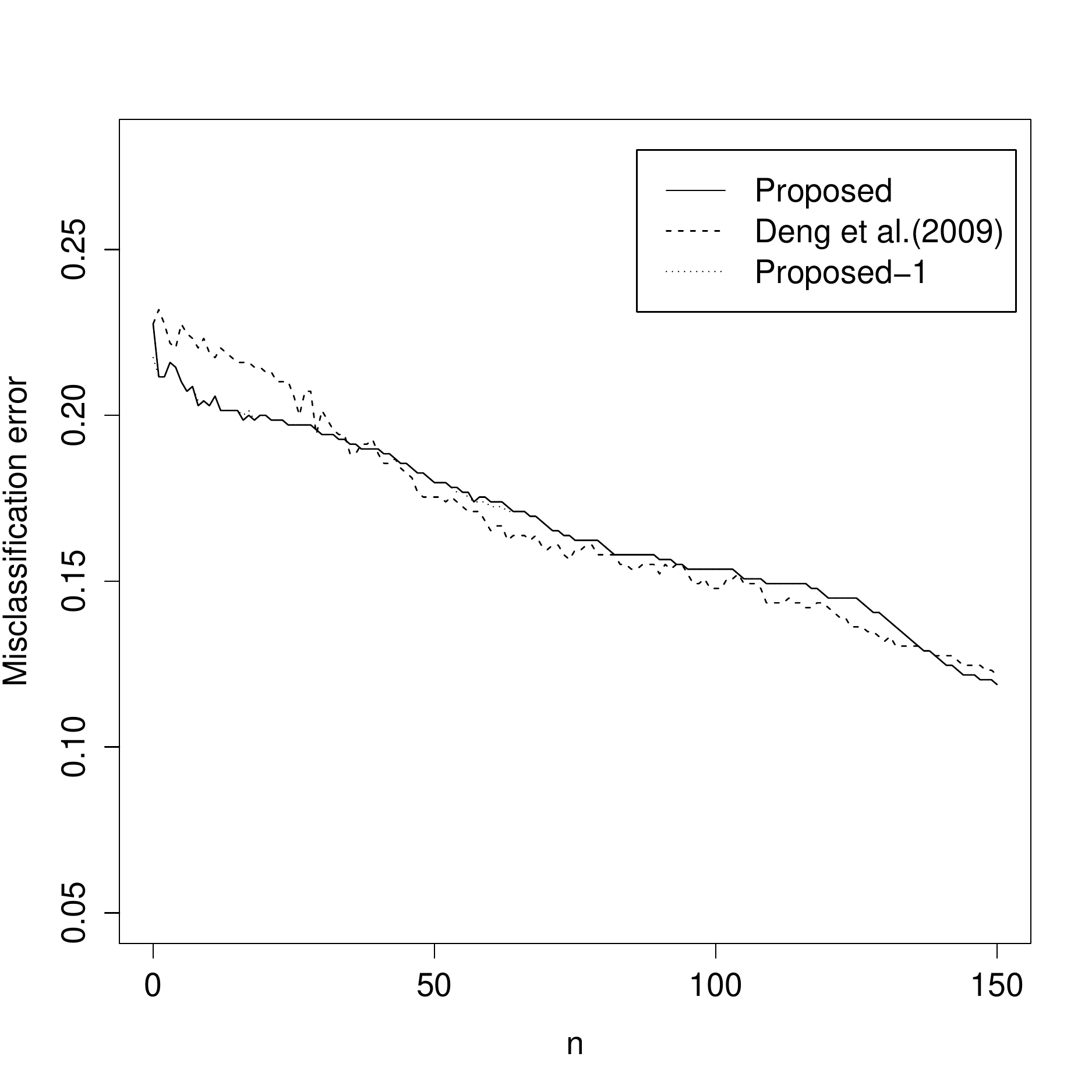}
	            \centering
	            %\caption{(a)}
	            %\label{fig:notification-sys:a}
	            \end{minipage}
	        }
           \subfloat[ BUPA data with $ n_0 = 30 $ ]{
	            \begin{minipage}[t]{ 0.5\linewidth }
	            \includegraphics[width=7cm,height=5cm]{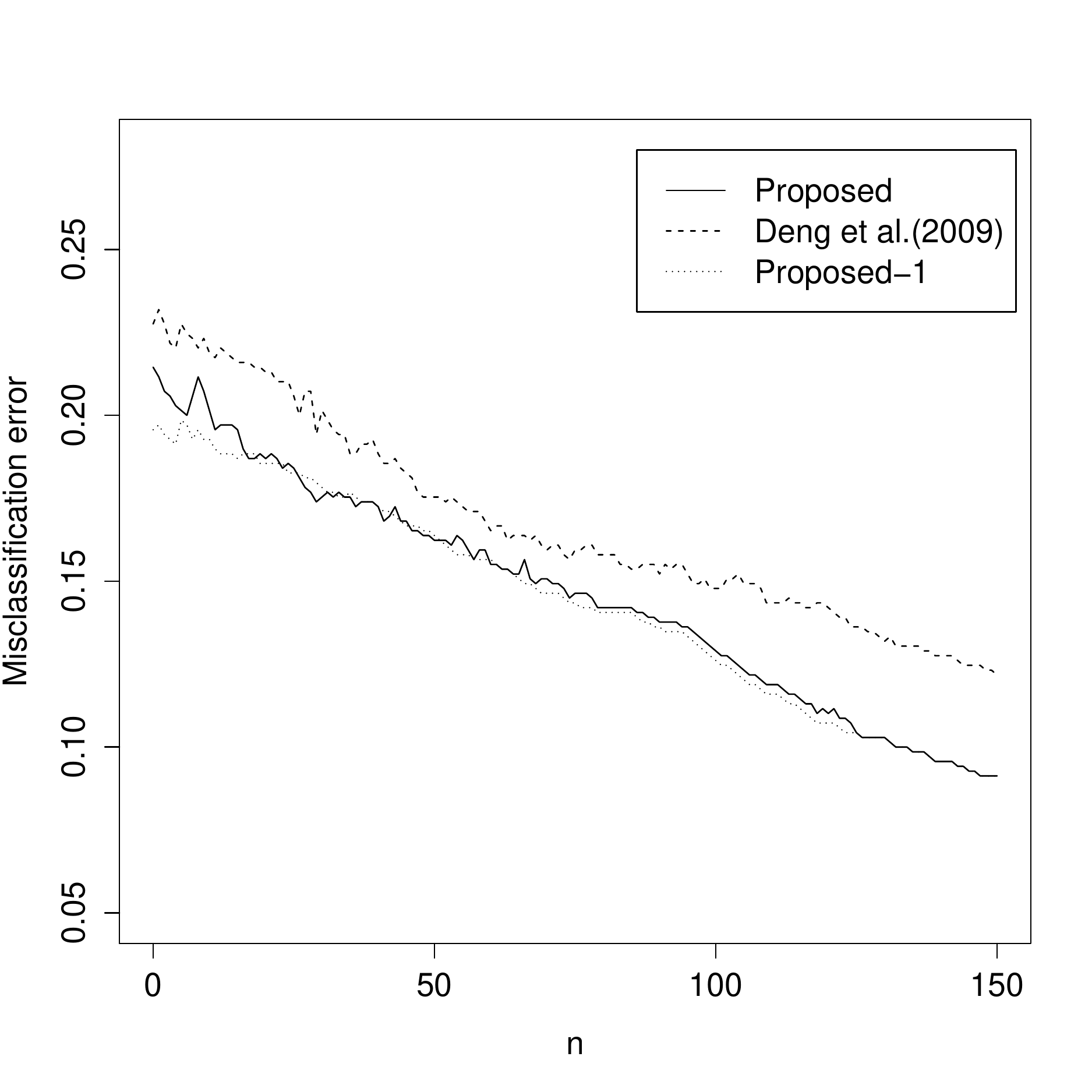}
	            \centering
	            %\caption{(a)}
	            %\label{fig:notification-sys:a}
	            \end{minipage}
	        } \\
            \subfloat[ WDBC with $n_0=0$ ]{
	            \begin{minipage}[t]{ 0.5\linewidth }
	            \centering
	            \includegraphics[width=7cm,height=5cm]{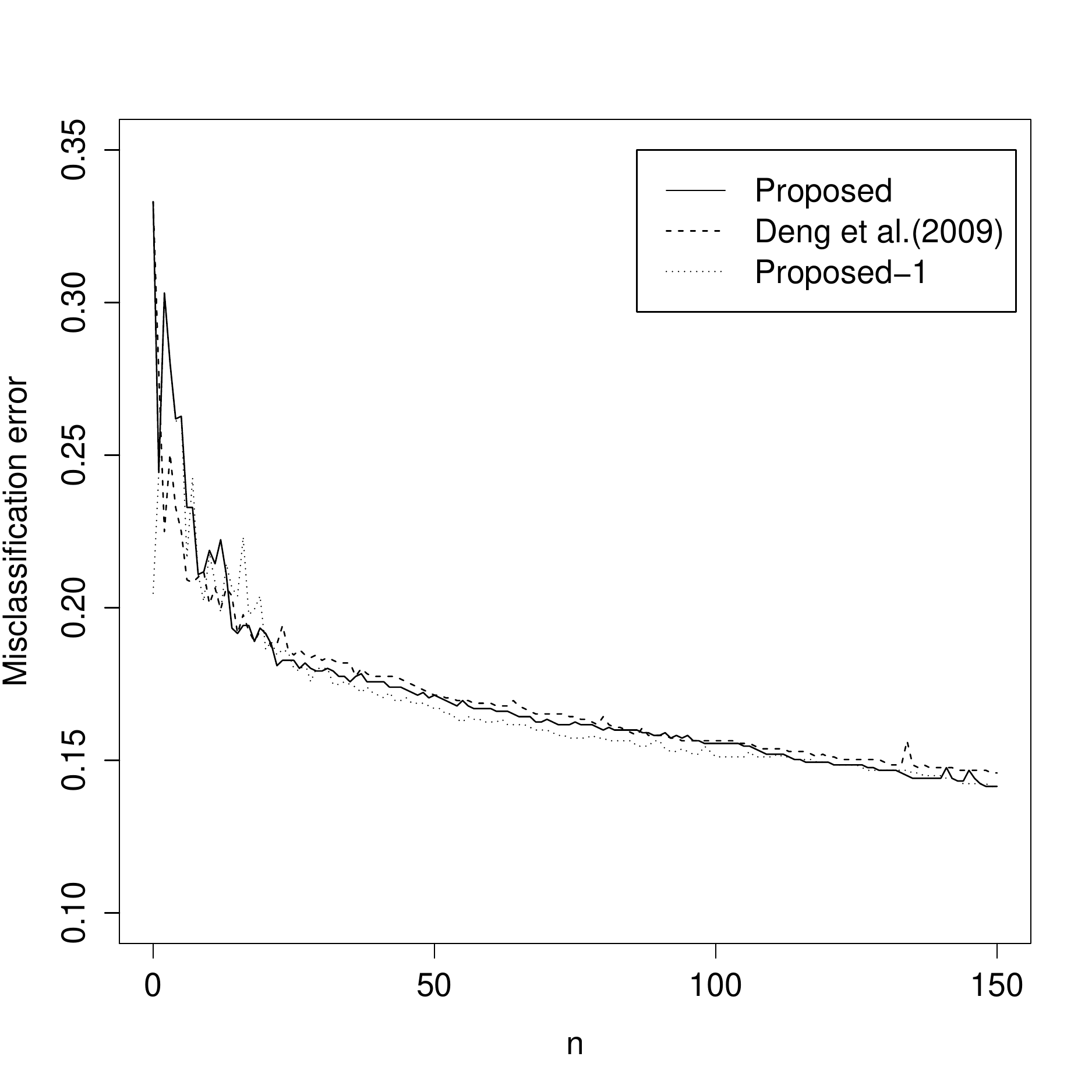}
	            %\caption{(b)}
	           % \label{fig:notification-sys:b}
	            \end{minipage}
            }
            \subfloat[ WDBC with $n_0=30$ ]{
	            \begin{minipage}[t]{ 0.5\linewidth }
	            \centering
	            \includegraphics[width=7cm,height=5cm]{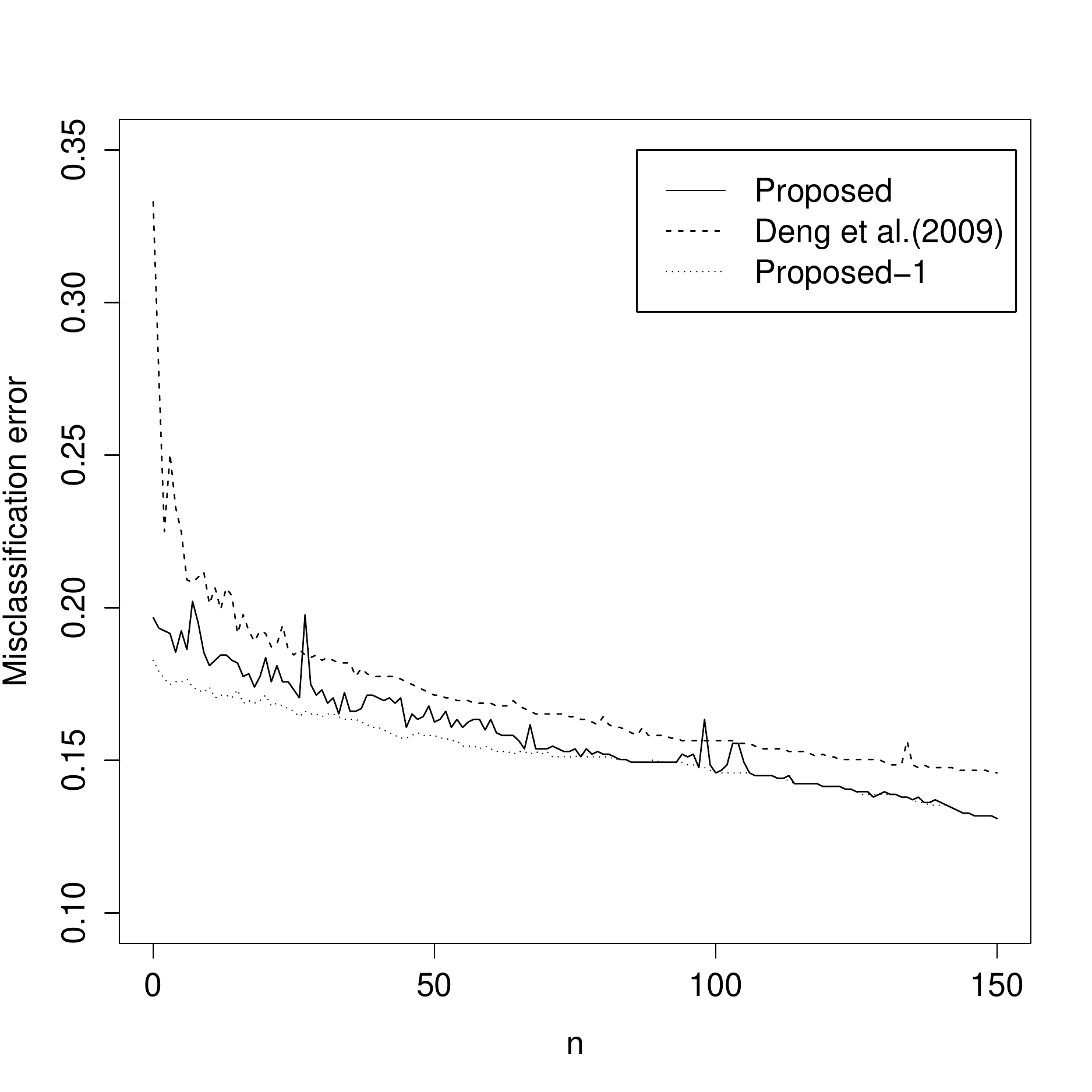}
	            %\caption{(b)}
	           % \label{fig:notification-sys:b}
	            \end{minipage}
            } \\
\caption{Misclassification curves when applying the proposed method with uncertainty measure with probability equal to 0.5 and adjusted cutting point using the proportion of sample sizes of two groups. Figures (a) and (c) show results with $n_0=0, and $Figures (b) and (d) are results with $n_0=30$. The black, blue and red misclassification curves are ALSD, the proposed method without and with adjusted for sample sizes, respectively. }
          \label{Fig:real-adj}
\end{figure}

\paragraph{Real Data Examples}
Similar results are obtained, when we apply both methods to BUPA and WDBC data sets. Figure \ref{Fig:real-adj} shows results of three different methods : ALSD, the propose method with and without sample sizes adjustment.  The last two methods are denoted as ``proposed-1'' and ``proposed'',  respectively. 

The ratio of sample sizes of two groups in BUPA data set is 0.58, which is close to 0.5.  Hence, the effect of uneven group sizes is not that obvious especially when $n_0=0$.  In WDBC data set, the ratio is  0.627, which is slightly far away from 0.5.  We can see from Figure \ref{Fig:real-adj} (c) that the misclassification curve of the proposed method with adjusted cutting point (proposed-1) becomes the best one when the number of the cumulated labeled subjects is larger than around 40. In  Figure \ref{Fig:real-adj} (d), it shows that with $n_0=30$, the proposed method with an adjusted cutting point (proposed-1) is the most stable one, among three methods from the very beginning.

The phenomenon of using two different parameters for uncertainty measure and cutting threshold in fact can be explained from a statistical decision theory viewpoint.  The details are discussed in Appendix A.

\section{Discussion}
\label{sec: Discu}
Active learning selects its own training samples in a sequential manner and requires fewer labeled instances from domain experts, and still achieves high classification performance. 
In this paper, we focus on a higher dimensional case and propose a new subject selection scheme that combines a Bayesian D-optimal design and an uncertainty sampling method. Thus, the proposed method inherits the advantage of methods of stochastic approximation and optimal design as suggested in \citet{Wu1985}. 
Because of using a Bayesian D-optimal design method, the active learning process is more stable in high dimensional cases even when the information matrix is nearly singular, and therefore will be more suitable for modern analysis with large data sets. 
In addition, we also demonstrate that with a small amount of labeled subjects are an initial training set, active learning process is more stable and efficient in both training time and the size of the labeled data.  For uneven group sizes case, we suggest to use separate parameters to control uncertainty sampling and adjust the cutting threshold for better performance.
From our numerical studies, we found that the uncertainty measure and the probability of a event might play different roles in an active learning process; especially  when the sizes of two groups are uneven.  We found that to use uncertainty measure at 0.5 and then adjust the boundary according to the proportion of group sizes as that in classical logistic regression models produces better results in our studies.

These types of methods are suitable for problems with large amount of unlabeled data available, and have great potential for analyzing  ``big data'' problems.   From practical viewpoints, to include one new subject at a time is not practical. Not only because of the computational efficiency, but also the operational complexity.  This is similar to the situation in clinical trials, where sampling in batch as in a group sequential procedure is usually preferred.  Moreover, to label an unclassified subject is not only time consuming, there is also some operational costs such as experts' charge and so on.  Hence, how to conduct an active learning process with a batch of updated subjects, and how to construct a classification rule with a satisfactory performance under a given budget constraint are important problems in both practical and theoretical viewpoints.

\section*{References}

%\bibliography{Active_learning}

\newpage
\section*{Appendix A: Statistical Decision Theory Viewpoint} %\label{app:a}

Let $P(Y=0)$ and $P(Y=1)$ be the prior probabilities of two groups, and $P(Y=0|\mathbf x)$ and  $P(Y=1|\mathbf x)$ are the corresponding posterior probabilities given $\mathbf x$.
In conventional statistical decision theory, when the prior probabilities, $P(Y=0)$ and $P(Y=1)$ is known and there is no other information available, the best decision rule $r(x)$, for any given subject, will be: $r(x)=0$, if $P(Y=0) > P(Y=1)$; $r(x)=1$, otherwise. (The decision function $r(x)=0$ denotes that the subject with explanatory variable $x$ is assigned to Class 0, and vice versa.)  When a logistic model is assumed, and suppose that the \textcolor{black}{ odds-ratio} satisfies that \textcolor{black}{$[P(Y=1|\mathbf x) / P(Y=1|\mathbf x)]=F(\mathbf x)$},  the problem becomes how to estimate the unknown $F$, and the decision rule will be made based on the posterior probabilities given observed $\mathbf x$; that is,  $r(x)=0$, if $P(Y=0|\mathbf x) > P(Y=1|\mathbf x)$, and $r(x)=1$, otherwise.
It follows from Bayes formulae, this decision rule is equivalent to
\begin{align}\label{eq:Bayes}
	r(x) =
	\begin{cases}
	0,  & \text{if}~ P(\mathbf x| Y=0) P(Y=0) > P(\mathbf x | Y=1)P(Y=1); \\
	1,   & \text{otherwise}.
	\end{cases}
\end{align}
That is,  when a logistic model is used in a classification problem, based on (\ref{eq:Bayes}) the prior probabilities of two groups are already considered.  Thus, it suffices to use $F(\mathbf x) = 0.5$ to measure the uncertainty.

Moreover,  let $c_0>0$ and $c_1>0$ be the misclassification costs of false positive and false negative errors, respectively. If we introduce these costs of misclassification into the decision rule, then Bayes decision rule becomes $r(x)=0$, if $P(\mathbf x| Y=0) P(Y=0)/P(\mathbf x | Y=1)P(Y=1) > c_1/c_0$; $r(x)=1$, otherwise. Because $c_0$ and $c_1$ can be treated as weights of two types of misclassification errors, we can assume that $c_0+c_1=1$, and the overall misclassification error becomes $c_0 P_{fp} + c_1 P_{fn}$, where $P_{fp}$ and $P_{fn}$ denote the false positive and false negative probabilities.  This weighted misclassification error can usually be estimated by $c_0 FP+ c_1 FN$ \citep[see][]{Webb2011}.
In fact, in \citet[page 975]{Deng2009}, they also measured the misclassification  using this formulae, which is $(\alpha FP+ (1-\alpha) FN)/N$ in their notations, where $FP$ and $FN$ are numbers of false positive and false negative results.  That is, the same parameter $\alpha$, in their paper, is used to measure the uncertainty and to adjust the weights of two different types of errors as well.
From the discussion above, it is reasonable to treat the uncertainty measure and weights of misclassification errors, separately.
When prior probabilities are known,  we can use them to adjust the cutting point in order to minimize the weighted misclassification errors, but not the uncertainty measure.

In practice, these probabilities are usually unknown, so it motivates us an interesting future study -- ``whether can we use the estimated ratio of sample sizes to adjusted the cutting point?'' Moreover, because  active learning processes are conducted sequentially.  It is naturally to ask whether we can apply a stopping rule to a learning process with a pre-fixed performance target.  All these issues are related sequential estimate of the event probability under adaptive sampling and the results will be reported elsewhere.

\section*{Appendix B:}
\paragraph {\textcolor{black}{Using $ \mathbf{\phi_{1}(d)}$ as an approximation to $\phi(\mathbf{d})$}} %\label{app:b}
\textcolor{black}{Here we discuss how we can use $ \mathbf{\phi_{1}(d)}$  in (\ref{model-5}) to approximate $\phi(\mathbf{d})$ in (\ref{model-4}). }
The Bayesian D-optimality criterion of \citet{CL1989} is
\[
 \begin{split}
 \phi(d) & = \boldsymbol{E}_{\boldsymbol\beta}\{ \log(|\mathbf{I}(\boldsymbol\beta;d)|)\}
          = \int \log(|\mathbf{I}(\boldsymbol\beta;d)|)\pi(\boldsymbol\beta)\, d\boldsymbol\beta  \\
         & \approx \int \log(|\mathbf{I}(\boldsymbol\beta;d)|)f(\boldsymbol\beta|Y)\, d\boldsymbol\beta
          = \mu
 \end{split}
\]

Here expectation  $\boldsymbol{E}_{\boldsymbol\beta}$ is taken with respect~to~a~prior~ distribution~for $\boldsymbol\beta$.
$\pi(\boldsymbol\beta)$ is~the~prior~distribution~on $\boldsymbol\beta$.
 $f(\boldsymbol\beta|Y) $ ~is~the~current~posterior~distribution~of $\boldsymbol\beta$.

According to the importance sampling approach discussed in  \citet{givens2012computational}, $ \mu $ can be written in the form
\[
   \mu = \frac{ \int\log(|\mathbf{I}(\boldsymbol\beta;d)|) \frac{f(\boldsymbol\beta|Y)}{\pi(\boldsymbol\beta)} \pi(\boldsymbol\beta)\, d\boldsymbol\beta }{ \int \frac{f(\boldsymbol\beta|Y)}{\pi(\boldsymbol\beta)}\pi(\boldsymbol\beta)\, d\boldsymbol\beta },
\]
where the prior $ \pi(\boldsymbol\beta) $ serves as the importance sampling distribution. Draw $ \boldsymbol{\beta}_1,\cdots, \boldsymbol{\beta}_M $ $ \mathrm{i.i.d} samples $ from $ \pi(\boldsymbol\beta) $ and then the estimator is
$
   \hat{\mu}_{\mathrm{IS}} = \sum_{u=1}^{M} w(\boldsymbol{\beta}_u) \log(|\mathbf{I}(\boldsymbol{\beta}_u;d)|),
$
where $ w(\boldsymbol{\beta}_u) = w^{*}(\boldsymbol{\beta}_{u}) / \sum_{i=1}^{M} w^{*}(\boldsymbol{\beta}_{i}) $ and $ w^{*}(\boldsymbol{\beta}_{u}) = f(\boldsymbol{\beta}_u|Y) / \pi(\boldsymbol\beta_u) $.  \\
Applying
\[
   f(\boldsymbol{\beta}|Y) \varpropto L(\boldsymbol\beta) \pi(\boldsymbol\beta)
   \Rightarrow \frac{ f(\boldsymbol\beta|Y) }{ \pi(\boldsymbol\beta) } \varpropto L(\boldsymbol\beta)
   \Rightarrow \frac{ f(\boldsymbol\beta|Y) }{ \pi(\boldsymbol\beta) } = aL(\boldsymbol\beta)
\]
where $ a $ is a constant, we obtain $ w^{*}(\boldsymbol{\beta}_u) = aL(\boldsymbol{\beta}_u) $. Therefore,
\[
   w(\boldsymbol{\beta}_u) = \frac{ aL(\boldsymbol{\beta}_u) }{ \sum_{i=1}^{M} aL(\boldsymbol{\beta}_i) }
                           = \frac{  L(\boldsymbol{\beta}_u) }{ \sum_{i=1}^{M}  L(\boldsymbol{\beta}_i) }
                           = r_u.
\]
Thus,
\[
   \phi_{1}(d) = \hat\phi(d) = \sum_{u=1}^{M} r_u \log(|\mathbf{I}(\boldsymbol{\beta}_u;d)|).
\]

\end{document}